\documentclass{JINST}  

\usepackage{graphicx}

   \title{Planck LFI flight model feed horns}

\author{
Fabrizio Villa$^a$\thanks{Corresponding author.}, 
Ocleto D'Arcangelo$^b$, 
Massimiliano Pecora$^c$, 
Lorenzo Figini$^b$, 
Renzo Nesti$^d$, 
Alessandro Simonetto$^b$,
Carlo Sozzi$^b$,
Maura Sandri$^a$,
Paola Battaglia$^c$,
Pietro Guzzi$^c$\thanks{Present address: R\&D Technology Development, Numoyx Italy s.r.l., via C. Olivetti 2, 20041 Agrate Brianza (Mi), Italy},
Marco Bersanelli$^e$,
Reginald C. Butler$^a$,
~and Nazzareno Mandolesi$^a$\\
\llap{$^a$}Istituto di Astrofisica Spaziale e Fisica Cosmica, INAF, \\
via P. Gobetti, 101 -- I40129 Bologna, Italy\\
\llap{$^b$}Istituto di Fisica del Plasma, CNR,\\
  via Roberto Cozzi, 53 -- I20125 Milano, Italy \\
\llap{$^c$}Thales Alenia Space Italia, Sede di Milano, \\
S.S. Padana Superiore, 290 -- I20090 Vimodrone, Italy \\
\llap{$^d$} Osservatorio Astrofisico di Arcetri, INAF, \\
Largo E. Fermi, 5 -- I50125 Florence, Italy \\
\llap{$^e$}Universit\`a degli Studi di Milano, \\
Via Celoria 16, 20133 Milano, Italy\\

  E-mail: \email{villa@iasfbo.inaf.it}}

\abstract{The Low Frequency Instrument is optically interfaced with the ESA Planck telescope through 11 corrugated feed horns each connected to the Radiometer Chain Assembly (RCA). This paper describes the design, the manufacturing and the testing  of the flight model feed horns. They have been designed to optimize the LFI optical interfaces taking into account the tight mechanical requirements imposed by the Planck focal plane layout. All the eleven units have been successfully tested and integrated with the Ortho Mode transducers.}


\begin{document}

%

\section{Introduction}
The Planck mission has been developed to produce a deep, full-sky imaging of cosmic microwave background (CMB) in both temperature and polarization. It incorporates an unprecedented combination of sensitivity, angular resolution and spectral range, spanning from centimeter to sub-millimeter wavelengths, by integrating two complementary cryogenic instruments in the focal plane of the Planck telescope: the Low Frequency Instrument (LFI) covering the frequency range between 30 and 70 GHz and the High Frequency Instrument (HFI) covering the range between 100 and 857 GHz. 

LFI is an array of 22 pseudo-correlation radiometers mounted on 11 Radiometer Chain Assemblies \cite{bersanelli2009}. Each chain is interfaced with the telescope through a corrugated horn antenna specially designed to fit the optical requirements \cite{sandri2009} and the interface with the ortho mode transducers (OMT) \cite{darcangelo2009}. 
This kind of feed shows excellent features like high beam symmetry and thus low cross polarization, low sidelobe level, low attenuation and low return loss, making it the preferred choice for high sensitivity space application. Moreover, since they are widely studied antennas, software tools have been developed to accurately predict their electromagnetic characteristics. For the above reasons they have been chosen for LFI since the start of the project. Even though the design and fabrication of corrugated horns was never considered an issue for LFI, the mechanical constraints of the focal plane unit and the need to optimize the optical interfaces, made the design phase non-trivial.
\begin{figure}[!h]
\begin{center}
\includegraphics[width=0.8\textwidth]{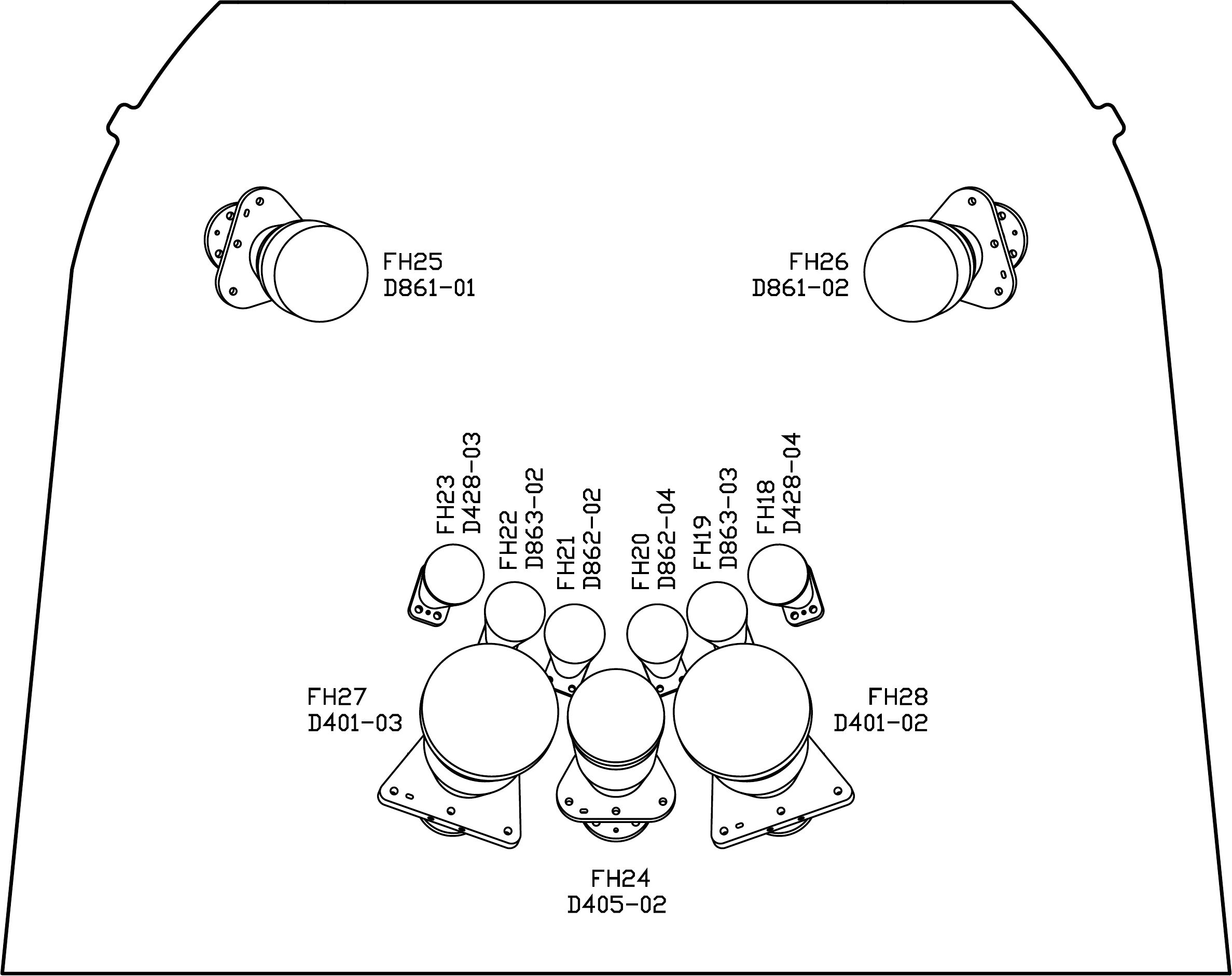}
\caption{The LFI feed horns front view in the focal plane. Each feed horn is labeled as FH$xx$, with $xx = 18 \to 28$. Different designs are labeled as D$yyy-zz$, according with the industrial design labeling: $yyy$ id the drawing reference number, $zz$ is the serial number of the flight unit.  Horn corrugations are not displayed. The external limit is the envelope of the focal plane unit.} \label{fig:focalplane}
\end{center}
\end{figure}

The LFI feed horns are the result of a very long process of design, manufacturing and testing.  In the late nineties, prototype demonstrators (PD) have been built and tested \cite{bersanelli1998}. Then the Elegant Bread Board (EBB) were developed: the requirements for these components were very similar to those of the flight model in order to identify any possible criticality, to determine the manufacturing technology and also the mechanical tolerances \cite{villa2002}. Successively four qualification model (QM), 
were built and tested \cite{darcangelo2005}. One horn at 30 GHz, one at 44 GHz, and two horns at 70 GHz,  demonstrating the possibility to reach the requested performance. Last, eleven flight model (FM) feeds were built and extensively tested based on six different electromagnetic designs because of the symmetry of the focal plane as shown in Fig \ref{fig:focalplane}. In particular the D401 and D405 were identical to the QM units, while at 70 GHz the horn designs were revised to improve as much as possible the overall LFI angular resolution.  A set of flight spare (FS), one per design, was also manufactured according to the overall LFI model philosophy. 

In chapter \ref{chapt:design} the electromagnetic design is described. Chapter \ref{chapt:phase_centre} reports the discussion about the phase center and its definition used in the case of LFI. Chapter \ref{chapt:dev_manuf} describes the development of the feed horns and the manufacturing solutions adopted. Qualification, with particular emphasis on electromagnetic horn testing is reported in chapter \ref{tests}. Conclusions are addressed in chapter \ref{chapt:conclusions}

\section{Electromagnetic Design}\label{chapt:design}
Dual profiled corrugated horns were selected due to their superior compactness, the possibility of optimizing the location of the phase center and shaping the pattern. In order to counterbalance the strong defocus due to the curved focal surface of the Planck telescope, the phase center position of the LFI horns, arranged around the HFI components, required an accurate control.
The phase center and the length were optimized for each horn depending on its position in the focal plane, while preventing possible blockage problems between horns \cite{ocleto2003}. The horn aperture diameter was mainly set to illuminate the telescope appropriately as a result of a trade off between angular resolution of the main beam and straylight rejection level due to sidelobes \cite{sandri2004, burigana2004}. 

\begin{table}
\caption{LFI feed horn design specifications. The edge taper is the taper used to optimize the optical response; the return loss and the cross polarization are the maximum design values over the whole bandwidth; The phase center location is defined as the distance between the horn flange and the focal surface of the telescope.   
}             
\label{tab:specs}      
\centering                          
\begin{tabular}{l r  r r r  r r }        
\hline                 
          & D401 & D405 & D861   & D428   & D862   & D863    \\
\hline
FH ID \dotfill&  27;28& 24 & 25;26  & 18;23 & 20;21& 19;22 \\
$\nu_0$ [GHz] \dotfill & $30$        & $44$    & $44$        & $70$       & $70$        & $70$   \\                
$\Delta\nu$ [GHz] \dotfill & $6$     & $8.8$    & $8.8$       & $14$         & $14$       & $14$  \\
Edge Taper [dB@$22^\circ$]\dotfill&  $30$ & $30$   & $30$  &$17$  & $17$ & $17$ \\
Return Loss [dB]\dotfill &  $-30$ & $-30$  &$-30$ & $-30$& $-30$&$-30$ \\
Cross Polarization [dB]\dotfill &  $-30$&$-30$  &$-30$ &$-30$ &$-30$ &$-30$ \\
Phase Center Location [mm]\dotfill & $119.60$ & $122.23$ & $83.83$ & $59.18$ & $66.13$ & $63.63$ \\
Total Length [mm] \dotfill & $156.12$ & $150.44$ & $133.04$ & $61.54$ & $68.49$ & $66.00$ \\ 
\hline                        
\end{tabular}
\end{table}
 
Starting from design specifications as in table \ref{tab:specs}, the design procedure is summarized as follows: 

\begin{itemize}
   \item The aperture diameter of the horns was determined to give the suitable co-polar width of its beam defined as the edge taper value \cite{sandri2009}
      \item The geometry of the corrugations was chosen to minimize the level of the cross polarization over the bandwidth. 
   \item The junction and the throat were designed to match the impedance of the smooth wall waveguide. 
  \item  The flare angle was chosen taking into account the length of the horn and the optimal beamwidth required to meet the telescope requirements.
\end{itemize}
Edge taper specifications are related to the simulation of performance of the feed-telescope system. An iteration process was followed, optimizing 
LFI's optical performances while considering the mechanical constraints at feed horn level. As far as feed horns are concerned, the main issue was to design of the 30 GHz one as short as possible. Moreover a challenge in the design was encountered for the design D861 due to the requirement on the phase center location that resulted well below the aperture plane (29.3 mm). At 70 GHz the circular smooth section length was minimized to reduce  the insertion loss in this part. 

The throat of the horn controls the  HE11 mode and the impedance matching, while the corrugation profile controls the HE12 mode and  the edge taper. Using a dual corrugated profile additional parameters were added to the optimization process and chosen to meet the design requirements, such as the position of the phase center and the radiation diagram, in a more compact structure with respect to the linear horn and with a smaller variation with  frequency \cite{gentili2001}. The process above led to a corrugation profile composed by a mixture of a sine--squared section \cite{olver1988}, apart for the D861 design, starting from the throat, and an exponential section near the aperture plane \cite{teniente2001, gentili2001}. 
The length of this last has a direct impact on the location of the phase center.
The analytical expression of the corrugation profile, $R(z)$, is the following
\begin{eqnarray}
R(z) =&& R_{th} + (R_s - R_{th}) \left[ (1-A)\frac{z}{L_s} + A \sin^\beta
\left(\frac{\pi}{2}\frac{z}{L_s} \right) \right] \\
&&0 \leq z  \leq L_s \nonumber
\end{eqnarray}
\noindent
in the sine section, and

\begin{eqnarray}
R(z) =&& R_s + e^{\alpha\left (z-L_s\right)}-1;\alpha = \frac{1}{L_e} ln(1+R_{ap}-R_s) \\
&&L_s \leq  z \leq L_e + L_s \nonumber
\end{eqnarray}
\noindent 
in the exponential region. Here, $R_{th}$ is the throat radius, $R_s$ is the sine squared region
end radius (or exponential region initial radius), $R_{ap}$ is the
aperture radius, $L_s$ is the sine squared region length and $L_e$
is the exponential region length. The parameter $A$ ($0 \le A \le 1$)
modulates the first region profile between linear and pure sine
squared type. The parameters $L_e/(L_e+L_s)$, $A$ and $R_s$ can be
used to control, as far as possible, the position and frequency
stability of the phase center and the compactness of the
structure. The feed horn parameters are reported in table \ref{table:designpar} 

\begin{table}
\caption{LFI feed horn parameters. $R_{th}$: Throat radius; $N_{s}$: Number of corrugation of the $sin^\beta$ section; 
$R_{sin}$: end radius of $sin^\beta$ section; $a$: tapering coefficient; $N_{e}$: number of corrugation of the exponential section; $R_{ap}$: final aperture diameter
}             
\label{table:designpar}      
\centering                          
\begin{tabular}{l r  r r r  r r }        
\hline              
          & D401 & D405 & D861   & D428   & D862   & D863    \\
\hline
$R_{th}$  & 0.49  &  0.49   & 0.49 & 0.5156 & 0.5156 & 0.5156  \\
$N_{s}$ &  26   &  35    & 32   & 34     & 40     & 39      \\
$R_{s}$ &  1.80 &  2.00    & 1.94 & 1.65   & 1.65   & 1.67    \\
$A$       &   0.80 &  0.75   & 0.67 & 1.0    & 1.0    & 1.0     \\
$N_{e}$ &  15  &  27    & 20   & 12     & 10     &   10    \\
$R_{ap}$  &  2.60 &   2.65 & 3.00 &  2.15  & 2.12   & 2.15    \\
$\beta$   & 2     &   2    &  3/2   &   2    &    2  &   2 \\
\hline                        
\end{tabular}
\end{table}
 
 \begin{figure}[!h]
\begin{center}
\includegraphics[width=1 \textwidth]{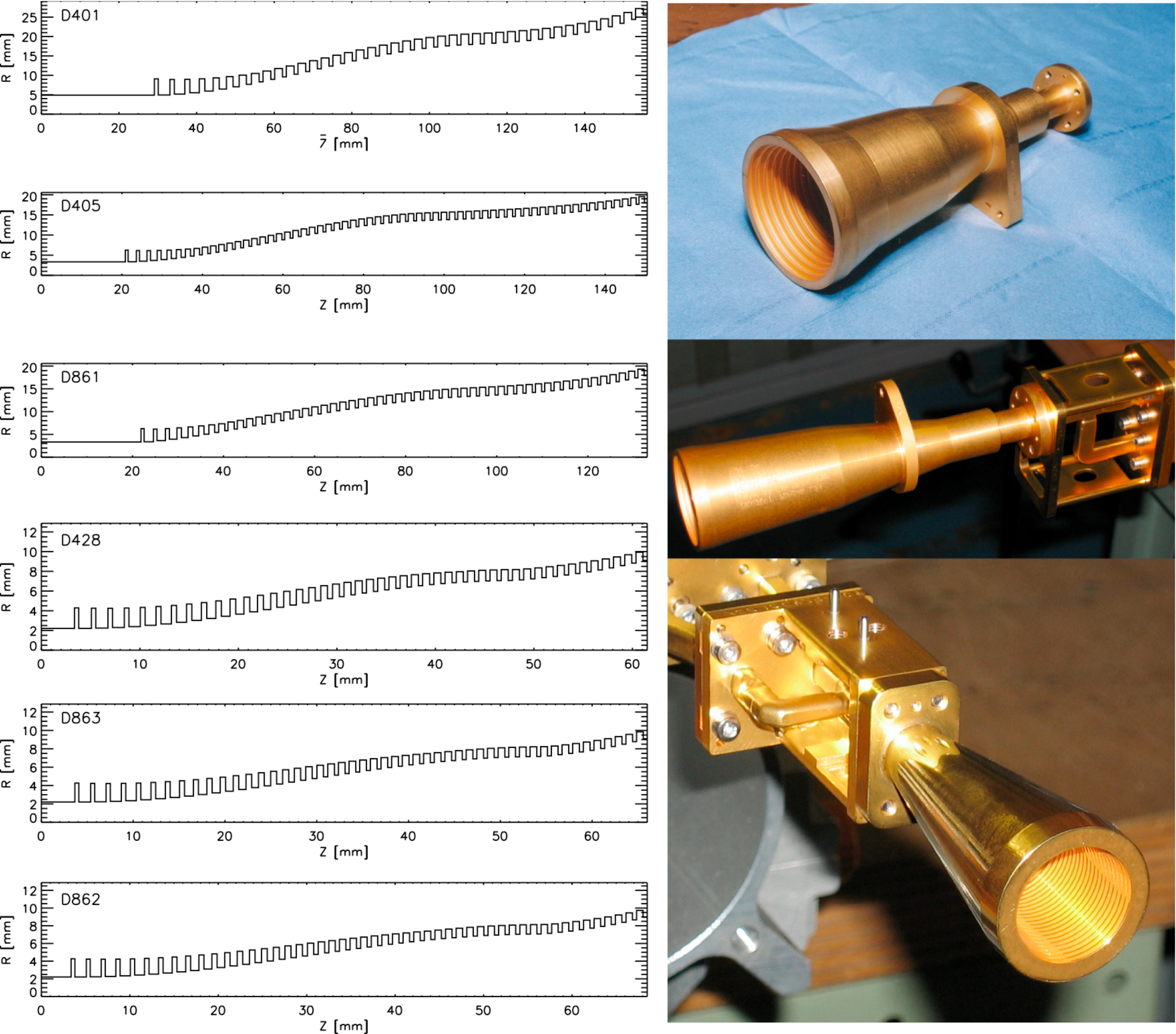}
\caption{{Left drawings: profile of corrugations; the range of each plot is the total lenght of the horn (X-axis) and the external radius (Y-axis). The picture on the right show the horns at 30, 44, and 70 GHz from top to bottom respectively.}} \label{fig:corrugations}
\end{center}
\end{figure}


\section{Horn Phase Center}\label{chapt:phase_centre}
Considerable effort was put, during the early development phases, in specifying the phase center location (PCL) inside feed horns. 
The PCL was defined as the distance between the feed / OMT interface plane (near the horn apex) and the intersection of the horn axis with the focal surface of the telescope. The design values, used for feed horn optimization,  are shown in Table \ref{tab:specs}.

As is well known, the far field phase front is spherical only for specially designed antennas, and no horn can make spherical wavefronts at finite distances. A rigorous analysis of the phase center concept for linear corrugated feed horns is given in \cite{wylde1993}, where several definitions, giving different results, are compared. 

\begin{figure}\
\centering
\includegraphics[width=0.8\textwidth]{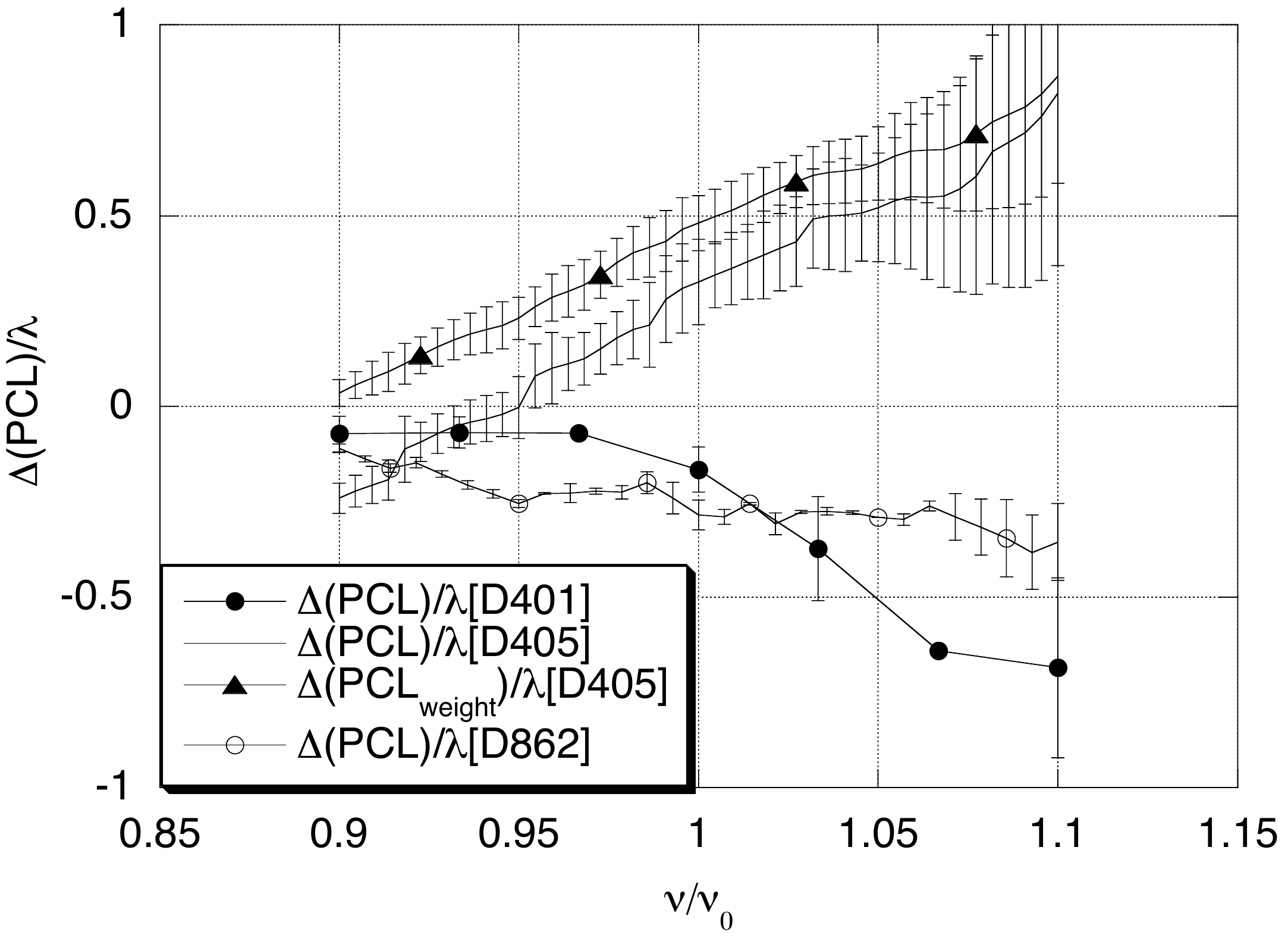}
\caption{Difference between estimated position of Phase Center and design goal. Data are normalized to wavelength, and are shown as a function of frequency, normalized to mid--band frequency. Three representative designs, one for each band, are shown. The phase center position is estimated using a best--fit to $-20$ dB with equal weights. Data for the $45^o$ plane are shown, with error bars representing the difference with E-- and H-- planes.
}
\label{fig:phaseCentre44}
\end{figure}

The horns were optimized using what in  the reference is named a \emph{least--squares--fit}  phase center definition at $-20$ dB. Using the radiated power as a weight does change the location of the computed phase center with respect to a fit with constant weight. But this variation is much smaller than the one intrinsically present across the operational bandwidth, as shown in figure \ref{fig:phaseCentre44}.

While this concept was used as a guidance for horn design, 
during acceptance tests it was soon realized that the definition of phase center used in optimization did not easily lend itself to measurements.  It was then decided that the most precise and appropriate way to measure compliance with the design parameters was computing the simulated radiated field at the exact distance where measurements were performed, and compare the phase patterns directly, without any reference to the ambiguous concept of phase center. The agreement was excellent, as shown in section \ref{patterns}.

\section{Development and Manufacturing}\label{chapt:dev_manuf}
\begin{figure}[!h]
\centering
\includegraphics[width=0.8\textwidth]{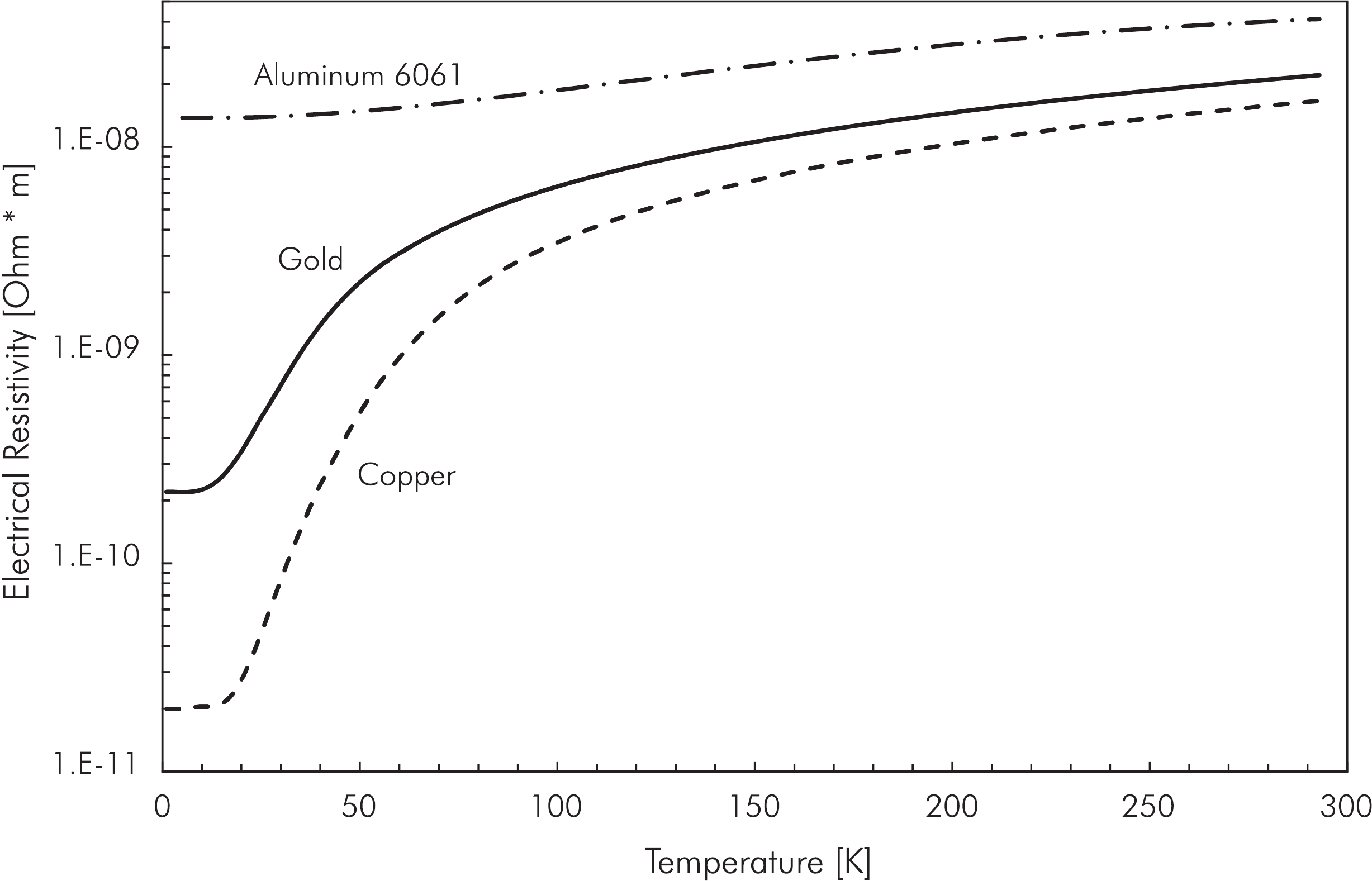}
\caption{Electrical resistivity of pure gold, pure copper, and aluminum 6061 \cite{clark1970},\cite{matula1979}.}
\label{fig:el_resistivity}
\end{figure}

Given the wide frequency band covered by the LFI horns, their dimensions and their internal corrugations are quite different. For this reason different manufacturing techniques were chosen: the 30 and 44 GHz were fabricated by FASTOR\footnote{FASTOR STAMPI s.r.l., Corso Cuneo, 41,10078 VENARIA (TO) - Italy,http://www.fastor.it}
using the electroerosion techinque. In fact the smaller depth of inner corrugations with respect to the throat diameter at these frequencies allowed to work directly an aluminum cylinder. The use of aluminum guaranteed the very low mass of the FH despite the larger dimensions. A gold film deposition was foreseen on the surfaces of the antennas, in order to protect it and to improve its  conductivity. Figure \ref{fig:el_resistivity} shows that the electrical resistivity is more than one order of magnitude lower for Gold than for Aluminum 6061, at 20 K (operational temperature).

On the other hand, it was not possible to use the same manufacturing technique at 70 GHz, given the small dimensions of corrugations. In this case the best solution to reproduce the geometry with high fidelity was given by the electroforming technique, which was successfully used since the first prototypes by CMi\footnote{Custom Microwave Inc., 24 Boston Ct. Longmont, CO 80501}.

\begin{figure}[!h]
\begin{center}
\includegraphics[width=1.\textwidth]{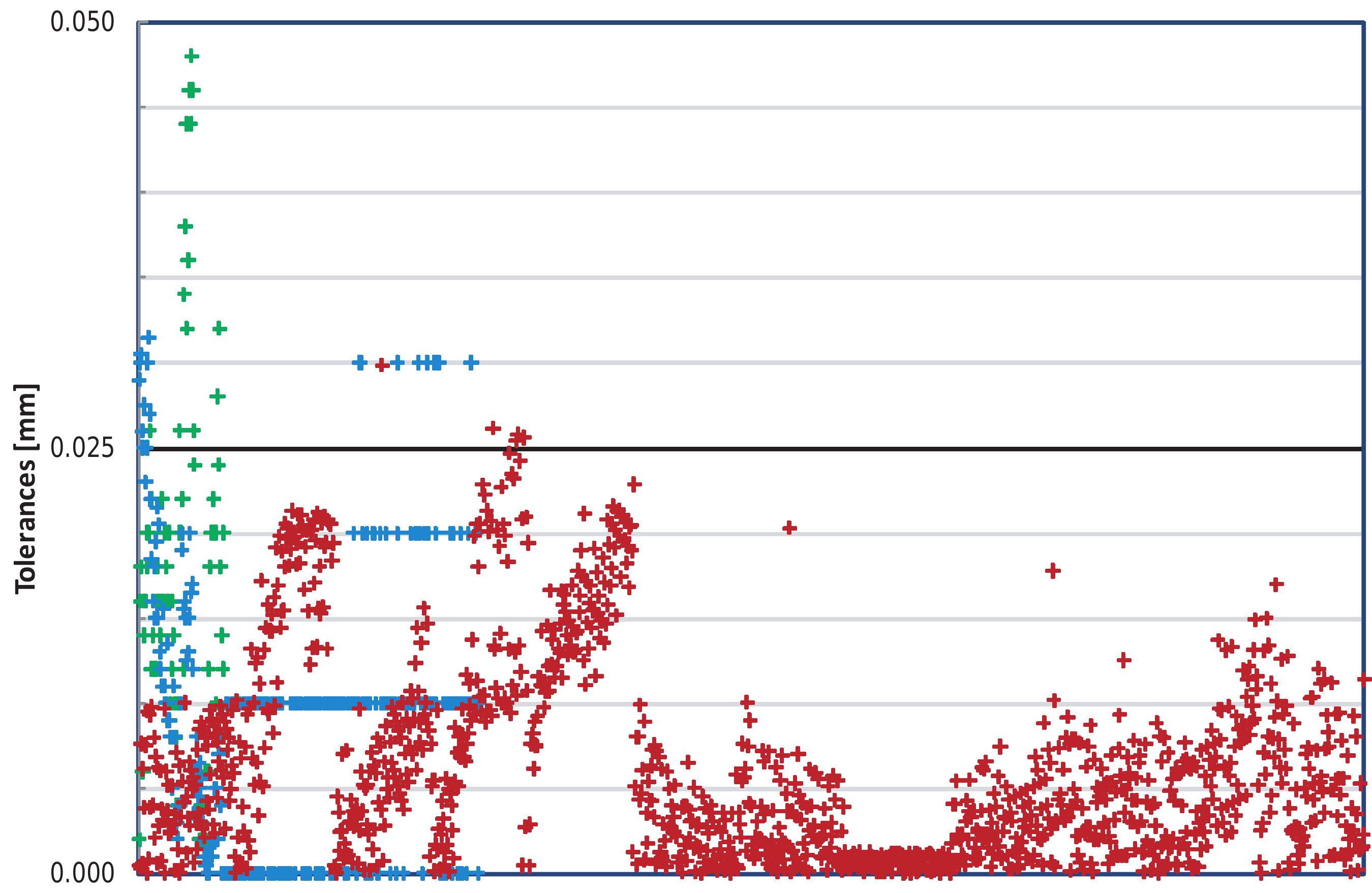}
\caption{The plot reports the deviation between the measured and the design corrugation dimensions. Both the diameter and the position of the corrugation are reported here. At 30 and 44 GHz the measurements were performed on the tools were used to manufacture the horns, while at 70 GHz measurements were performed on the aluminum master before the electroforming process. The green, blue, and red crosses are referred to the 30 GHz, 44 GHz, and 70 GHz feed horn respectively. The abscissa does not have physical meaning since it is just for plotting reasons. The solid horizontal line at 0.025mm represents the design tolerance. The points outside tolerances were evaluated case by case, although no RF simulations have been specifically performed to investigate the \emph{out-of-specification} points.}
\label{master}
\end{center}
\end{figure}

The typical manufacturing tolerance of the corrugation profiles was set to $0.025$ mm. 
The plots of Figure \ref{master} report the deviation of the corrugation with respect to the nominal dimensions. It was found that some of the corrugations were outside specifications; however it was verified case by case that this was not representing a problem concerning the electromagnetic performances. 

Anti-cocking custom flanges were designed to assure the correct mating with the OMT. At 30 GHz and 44 GHz the flanges were based on the UG383/U but with metric thread; at 70 GHz, due to mechanical constraints, the flange was rectangular instead of circular. 

\begin{figure}[!h]
\begin{center}
\includegraphics[width=1.0\textwidth]{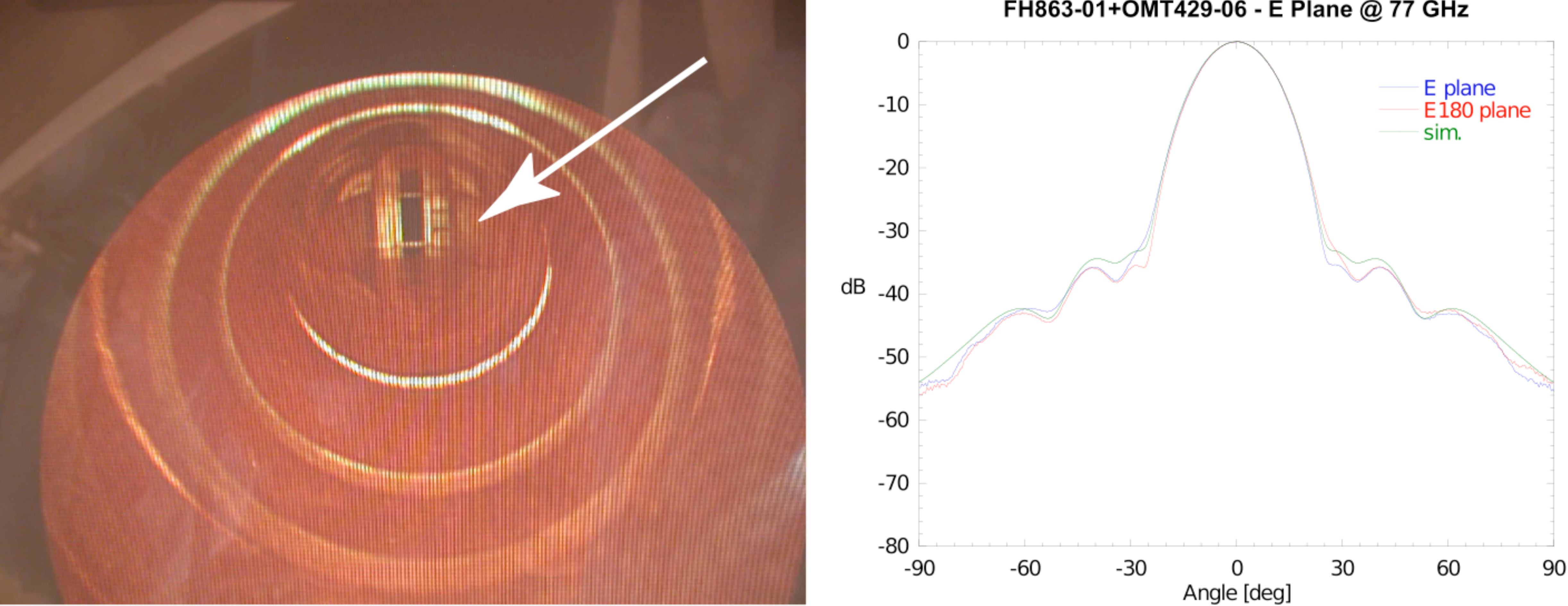}
\caption{The left panel shows the misalignment (indicated with the arrow) as photographed with the endoscope inside the horn at the OMT interface leve; the amount of misalignment was evaluated at about 0.14 mm as expected form pin/hole mechanical tolerances. Misalignment creates an asymmetry in the radiation pattern.  As shown in the right panel, the effect is clearly visible on the power radiation pattern at 77GHz, the highest frequency of the V-band horn (blue line) and in the pattern obtained by rotating the horn at 180 degrees around its axis (red line). The same pattern simulated  does not present asymmetries (green line).}

\label{fig:misalignment}
\end{center}
\end{figure}

To align the feed horns with the OMTs, alignment pins were used. The pin/hole mechanical coupling was standard (a pin diameter of $1.56$mm matched with a $1.7$ hole diameter) but inadequate to avoid effects on the electromagnetic performances of the feed/OMT assemblies. In fact,  the misalignment was discovered on RF pattern measurements on feed/OMT integrated units at 70 GHz for which asymmetries in the amplitude radiation patterns were found as reported in figure \ref{fig:misalignment}. After this discovery the integration and qualification sequence was updated and the assembly was integrated with the use of an optical endoscope. Moreover the baseline RF qualification procedure after vibration tests was changed to take into account the sensitivity to misaligment of the radiation pattern (see chapter \ref{vibration_test}).

\section{Testing and qualification }\label{tests}

Each FM FH went trough a very detailed electromagnetic qualification: measurements were performed using a Vector Network Analyzer (VNA)\footnote{AB millimetre, 52 Rue Lhomond, Paris, France http://www.abmillimetre.com} in a 7x5x4m anechoic chamber at CNR/IFP (Milano). The Horn Under Test (HUT) was placed on a manual XY translation stage, perpendicular to the rotation axis, mounted on a VNA-controlled rotary platform. The phase and amplitude patterns were acquired with the rotation axis passing through the aperture of the HUT. The alignment was aimed at getting a phase pattern as symmetric as possible in the central region. All the stands and the FH body were covered with Eccosorb$^{TM}$ in order to further reduce reflections that can have an impact in the far sidelobe region. 
A standard gain horn was used as fixed receiver at a distance of about 2 meters from the rotating source.  
A minimum of 40 beam patterns was measured for each FH. For each plane, the patterns are measured at five fixed frequencies inside the band (the central frequency and $\pm$5\%  $\pm$10\%  of this frequency). Then, after integration between the FH and OMT, all the beam patterns are measured once again at 70 GHz and only on the  principal planes (E/H) at 30 and 44 GHz, since mechanical constraints do not allow the measurement in the other polarizations for these lower frequencies. Also the Return Loss of all FHs was measured over the largest possible frequency band, in a one port reflectometry configuration, connecting the FH to the microwave circuit with a circular to rectangular transition, always using the VNA.

\subsection{Amplitude and phase patterns}\label{patterns}
\begin{figure}[!h]
\begin{center}
\includegraphics[width=0.48 \textwidth]{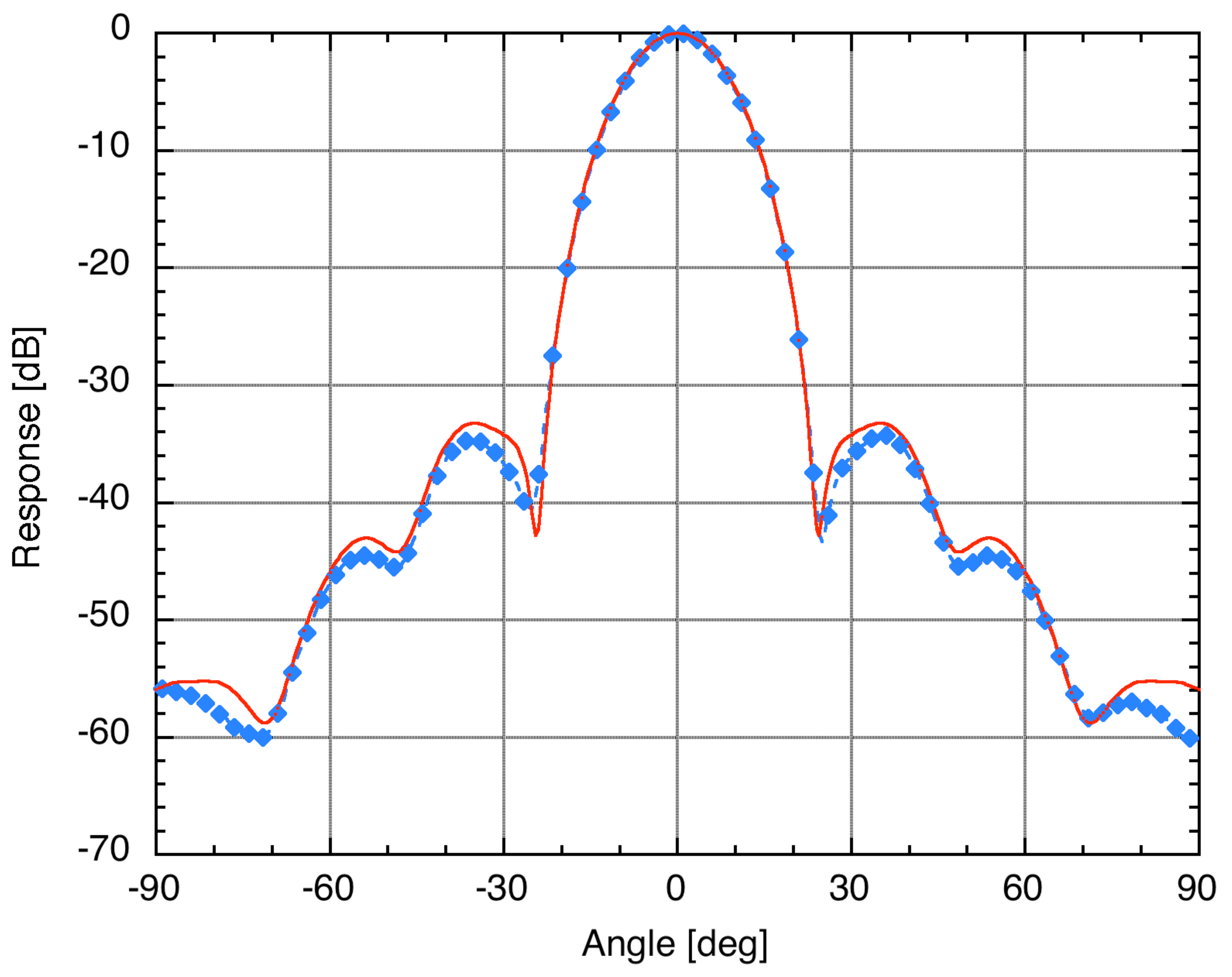}
\hfill
\includegraphics[width=0.48  \textwidth]{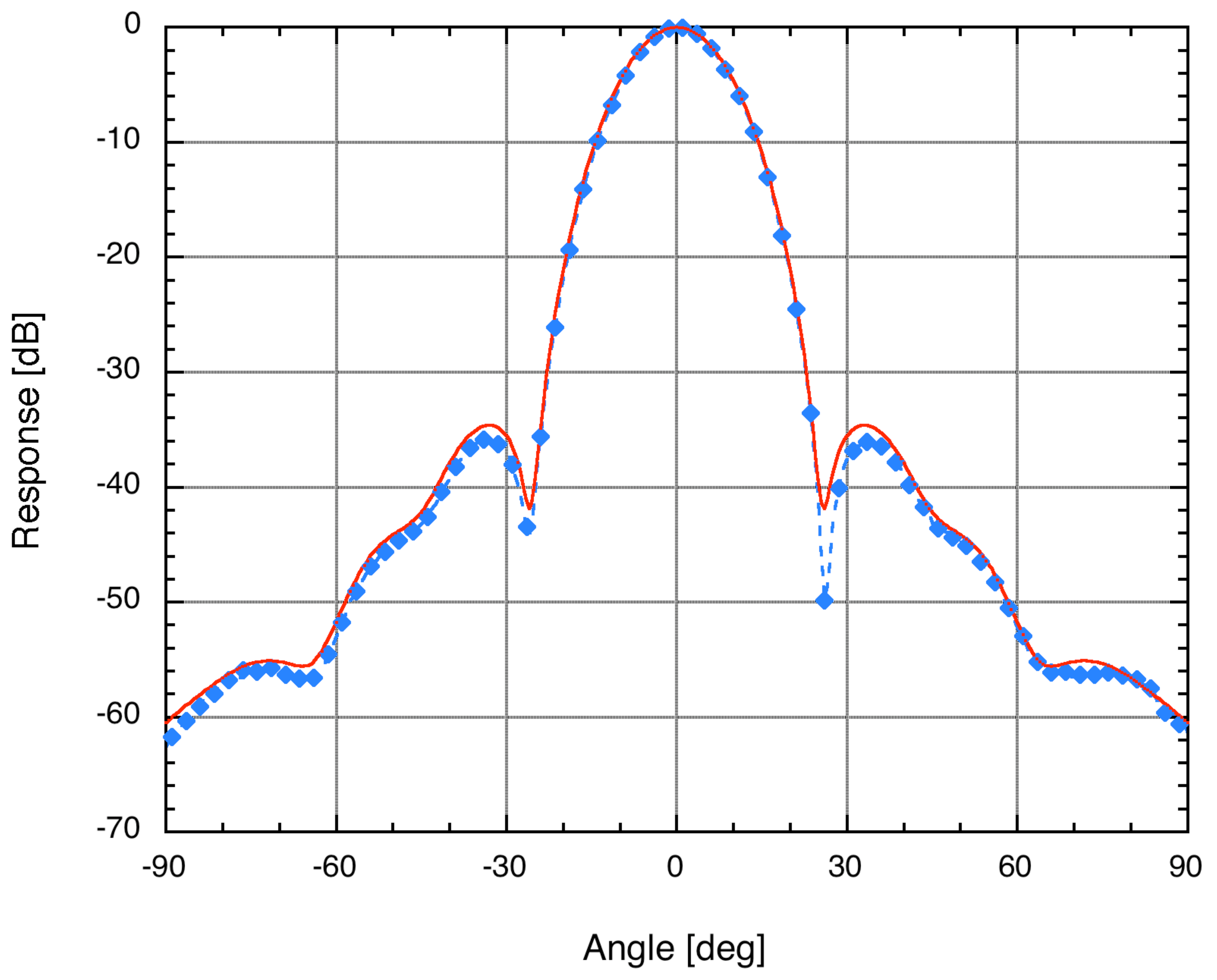}
\caption{E (left) and H (right) patterns measured at 30 GHz (symbols) compared with the simulated ones (solid lines).}
\label{EH30}
\end{center}
\end{figure}

\begin{figure}[!h]
\begin{center}
\includegraphics[width=0.48 \textwidth]{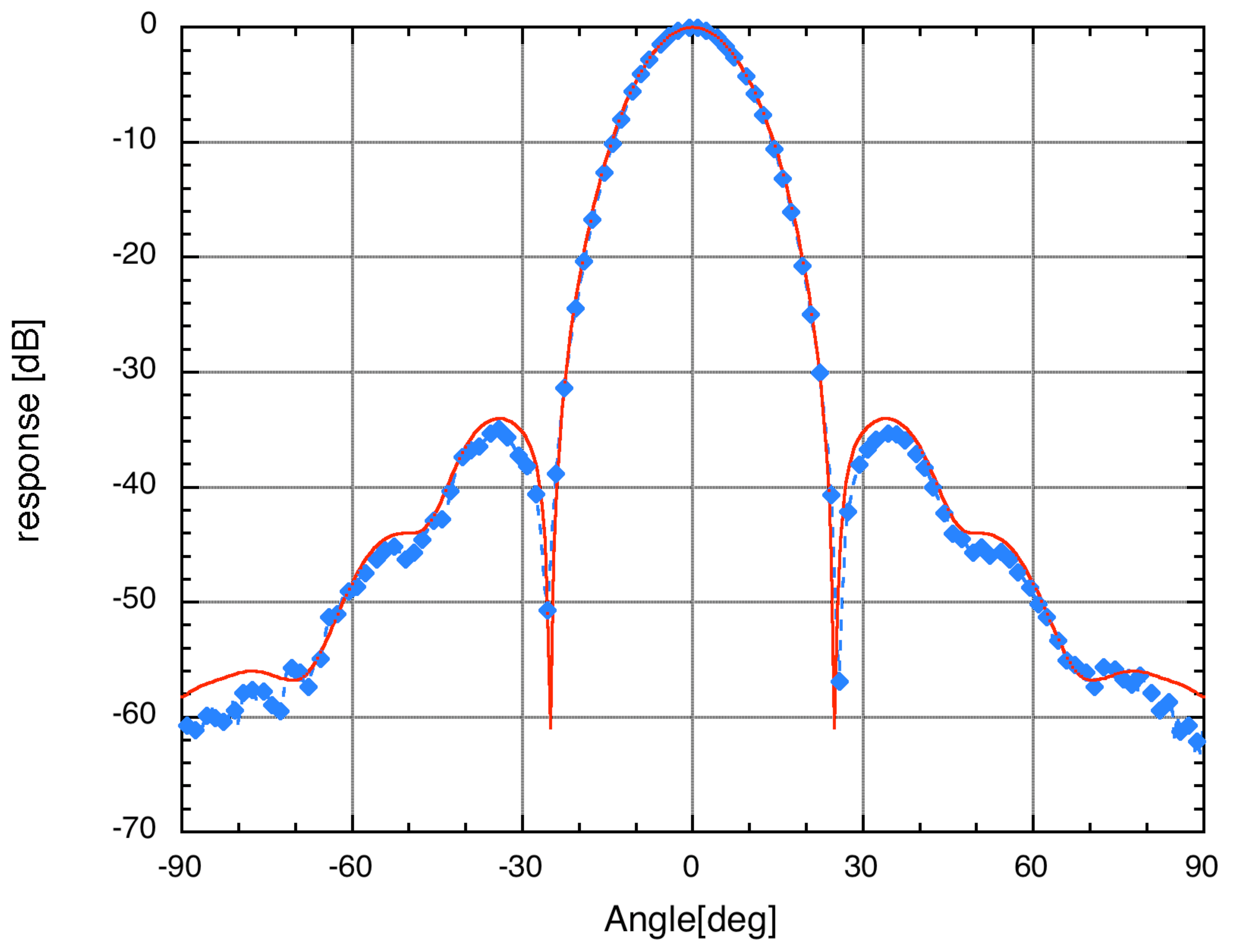}
\hfill
\includegraphics[width=0.48  \textwidth]{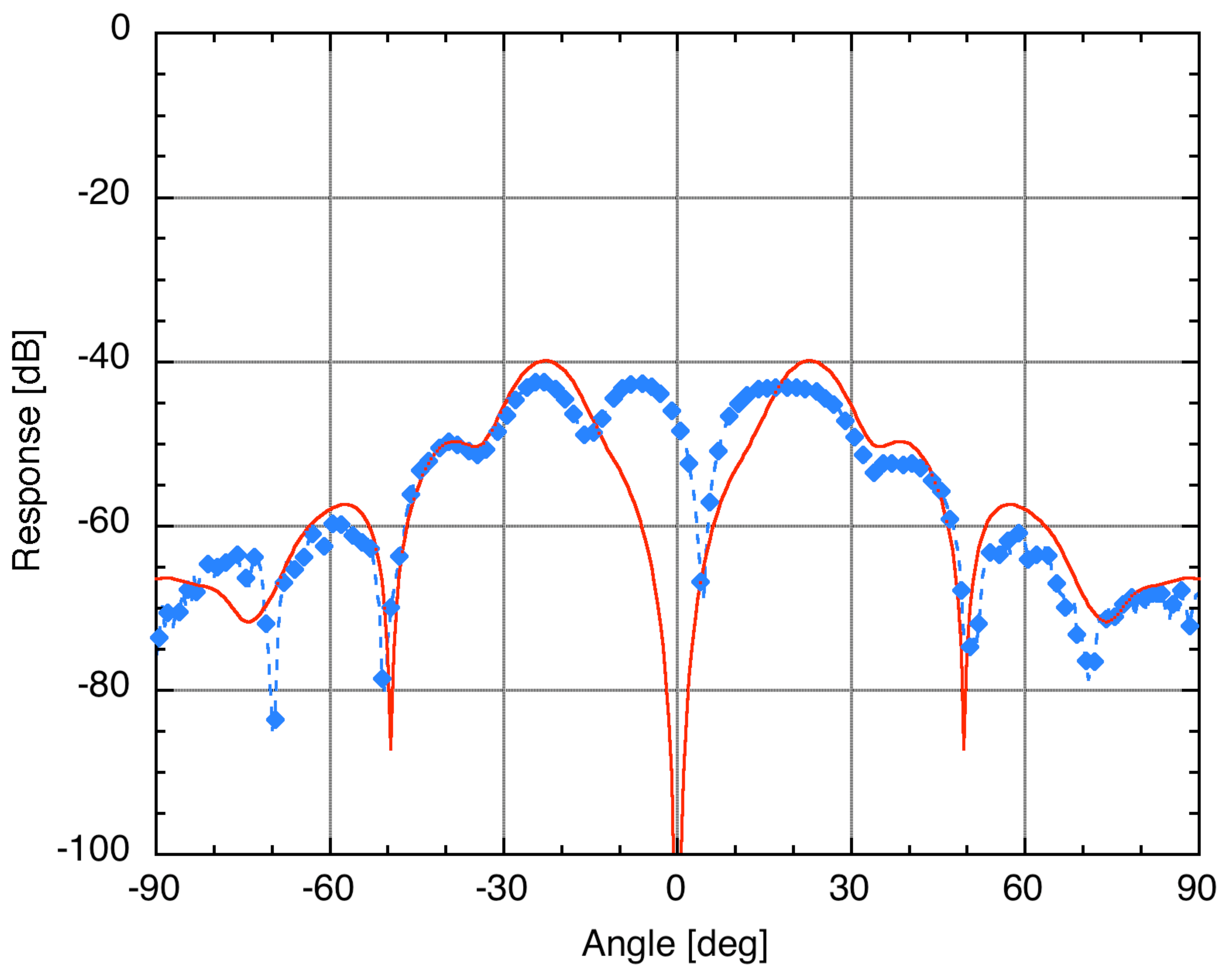}
\caption{45 degree plane co polar (left) and cross polar (right), measured (symbols) and simulated (solid line) patterns at 30 GHz.}
\label{4530}
\end{center}
\end{figure}

\begin{figure}[!h]
\begin{center}
\includegraphics[width=0.48 \textwidth]{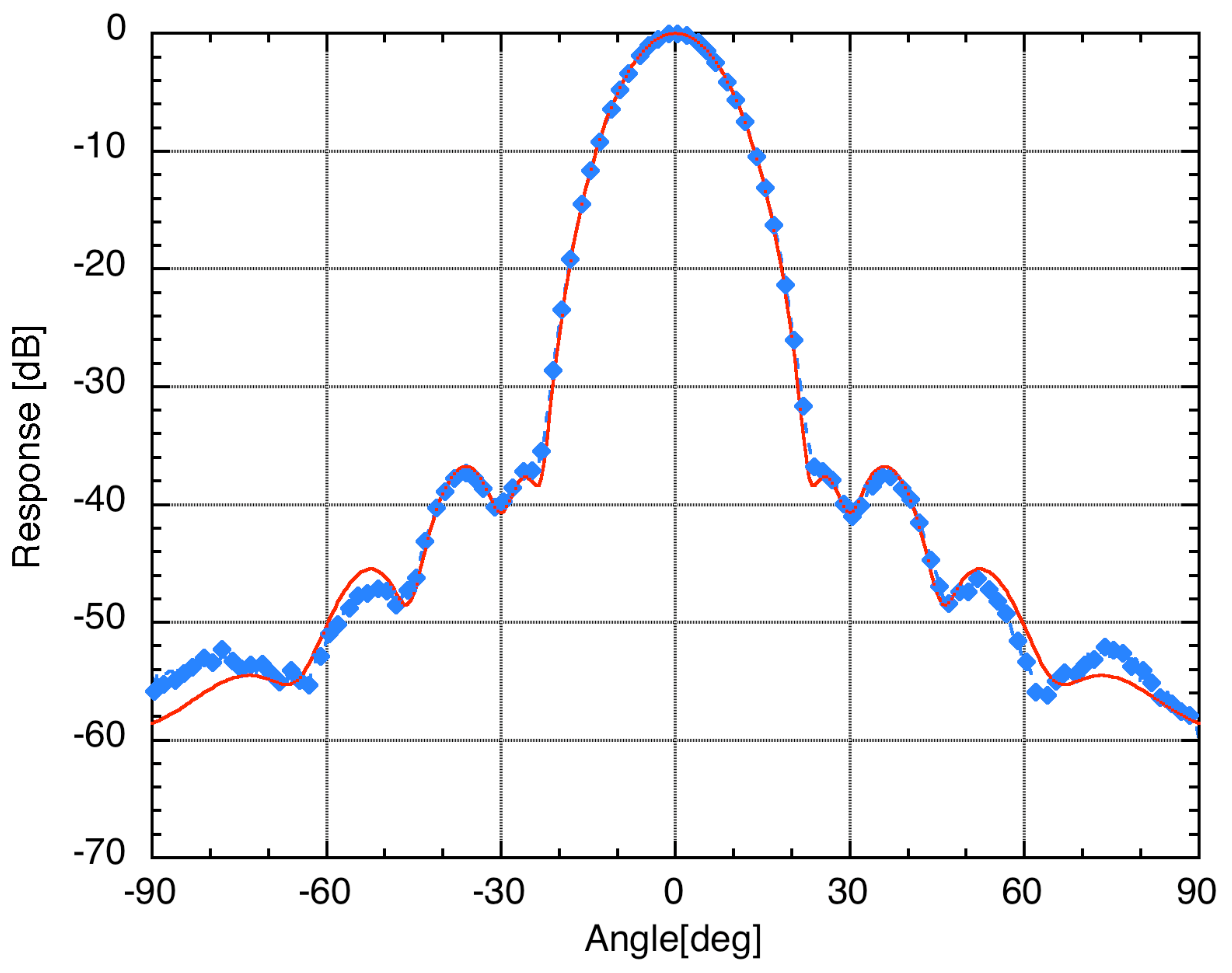}
\hfill
\includegraphics[width=0.48  \textwidth]{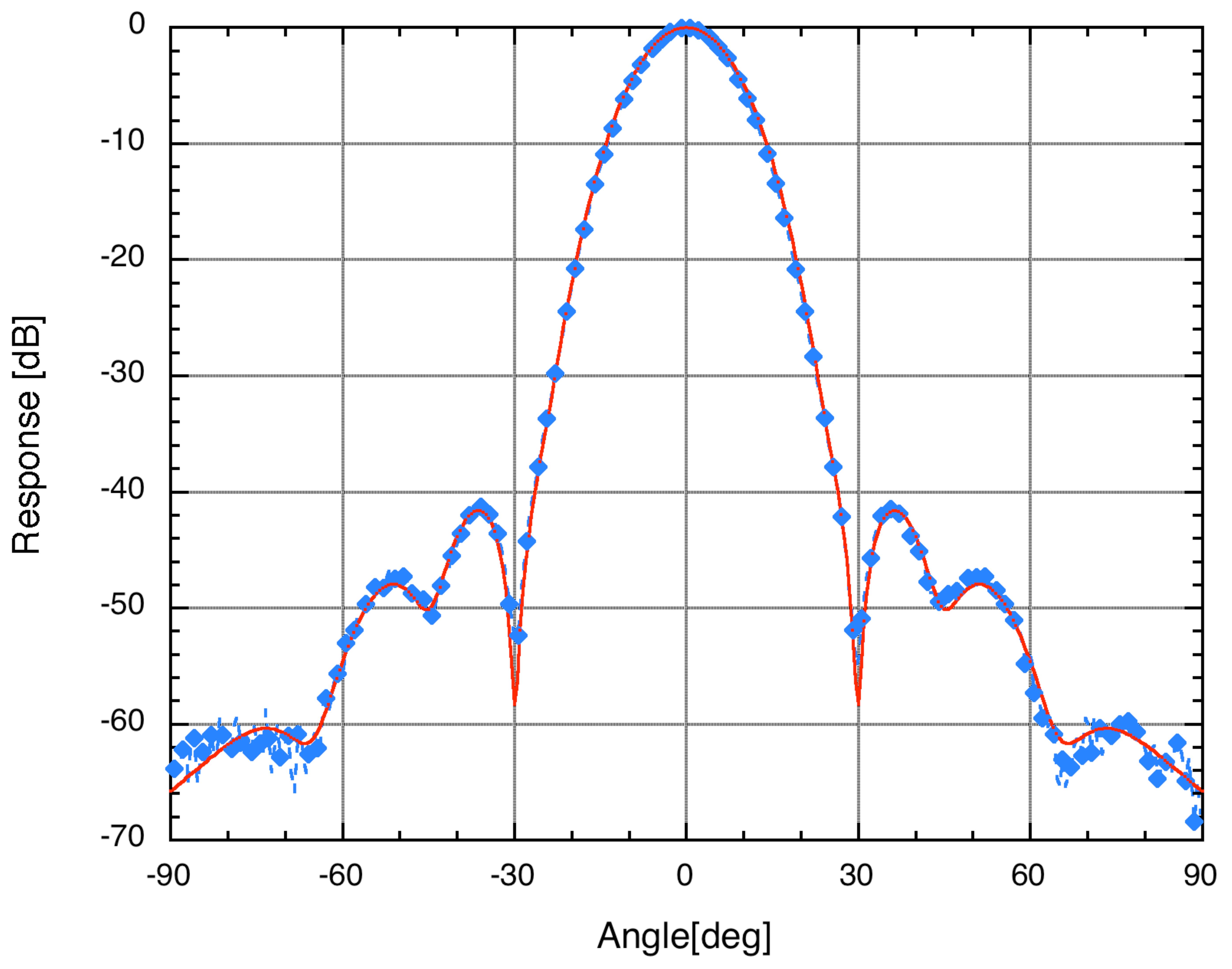}
\caption{E (left) and H (right) patterns measured  at 44 GHz (symbols) compared with the simulated ones (solid lines) }
\label{EH44}
\end{center}
\end{figure}

\begin{figure}[!h]
\begin{center}
\includegraphics[width=0.48 \textwidth]{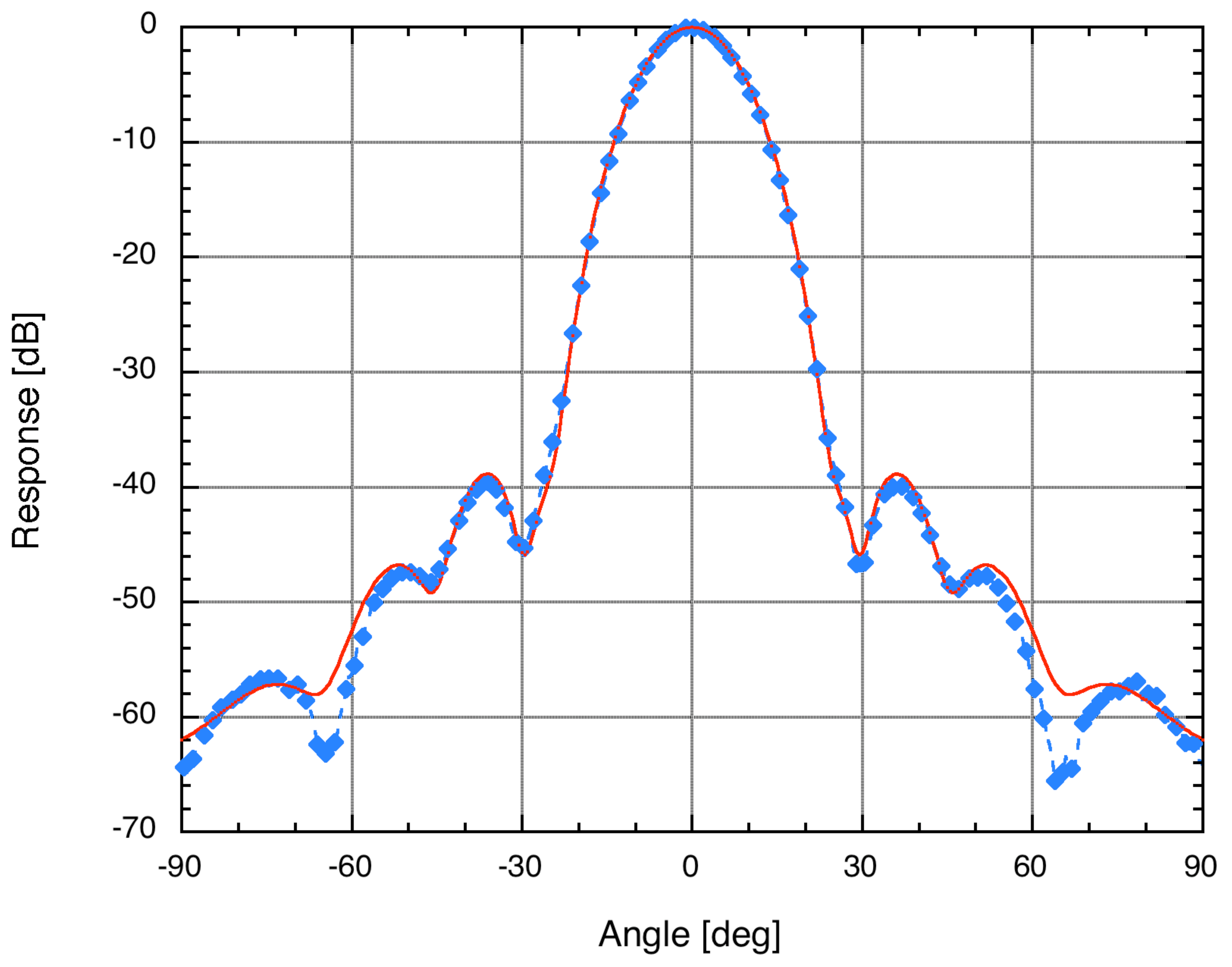}
\hfill
\includegraphics[width=0.48  \textwidth]{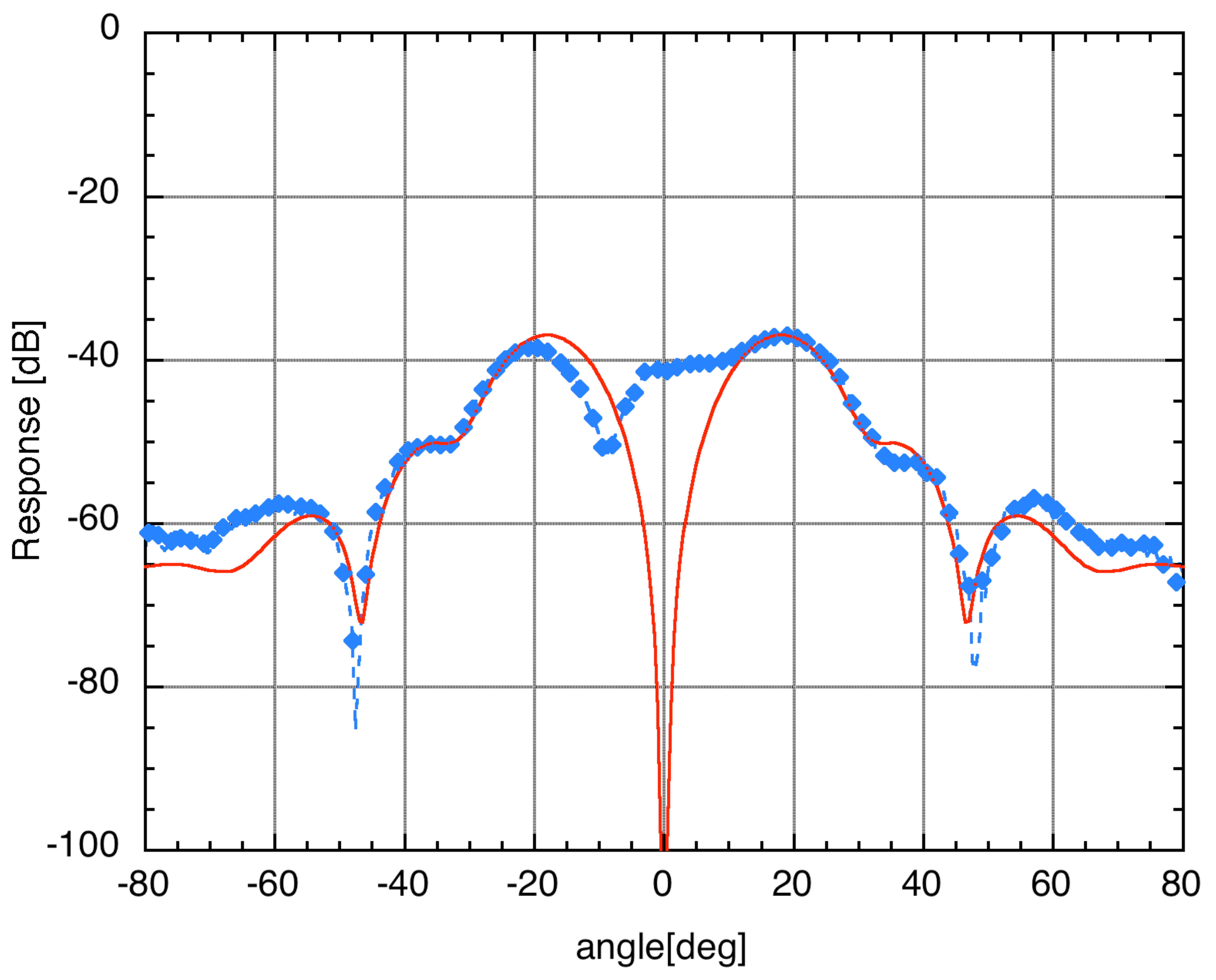}
\caption{45 degree plane co polar (left) and cross polar (right), measured (symbols) and simulated (solid line) patterns at 44 GHz.}
\label{4544}
\end{center}
\end{figure}

\begin{figure}[!h]
\begin{center}
\includegraphics[width=0.48 \textwidth]{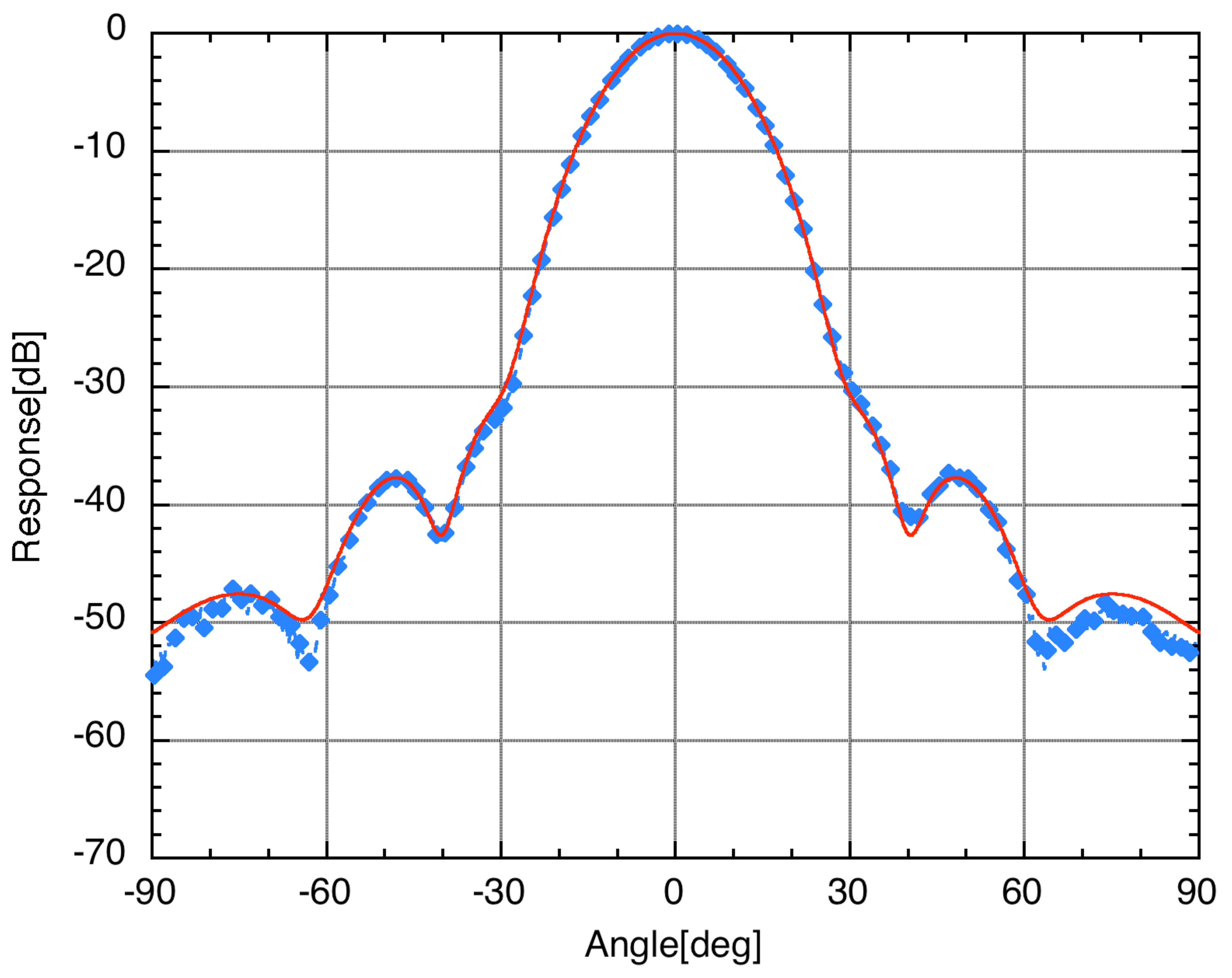}
\hfill
\includegraphics[width=0.48  \textwidth]{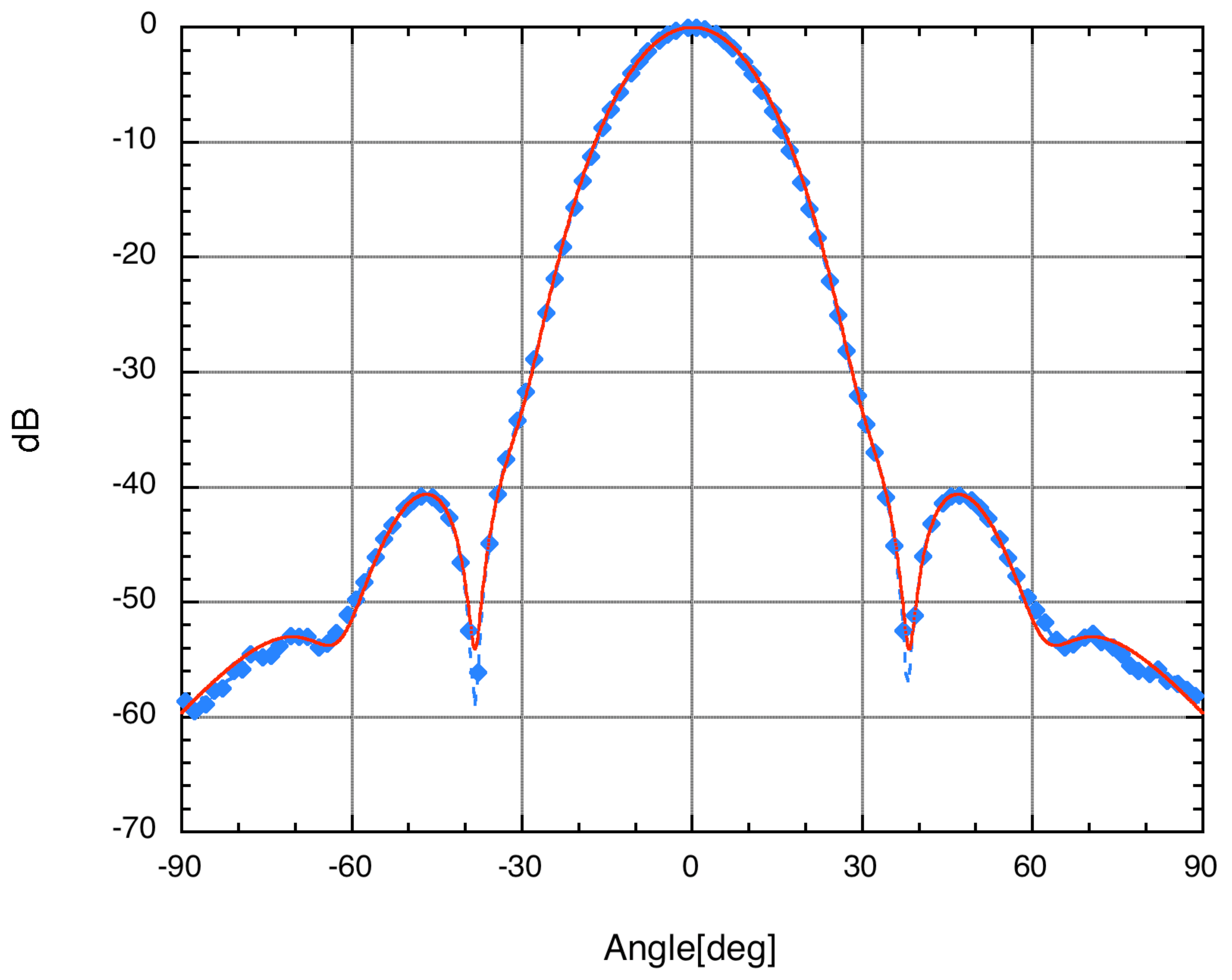}
\caption{E (left) and H (right) patterns measured at 70 GHz (symbols) compared with the simulated ones (solid lines).}
\label{EH70}
\end{center}
\end{figure}

\begin{figure}[!h]
\begin{center}
\includegraphics[width=0.48 \textwidth]{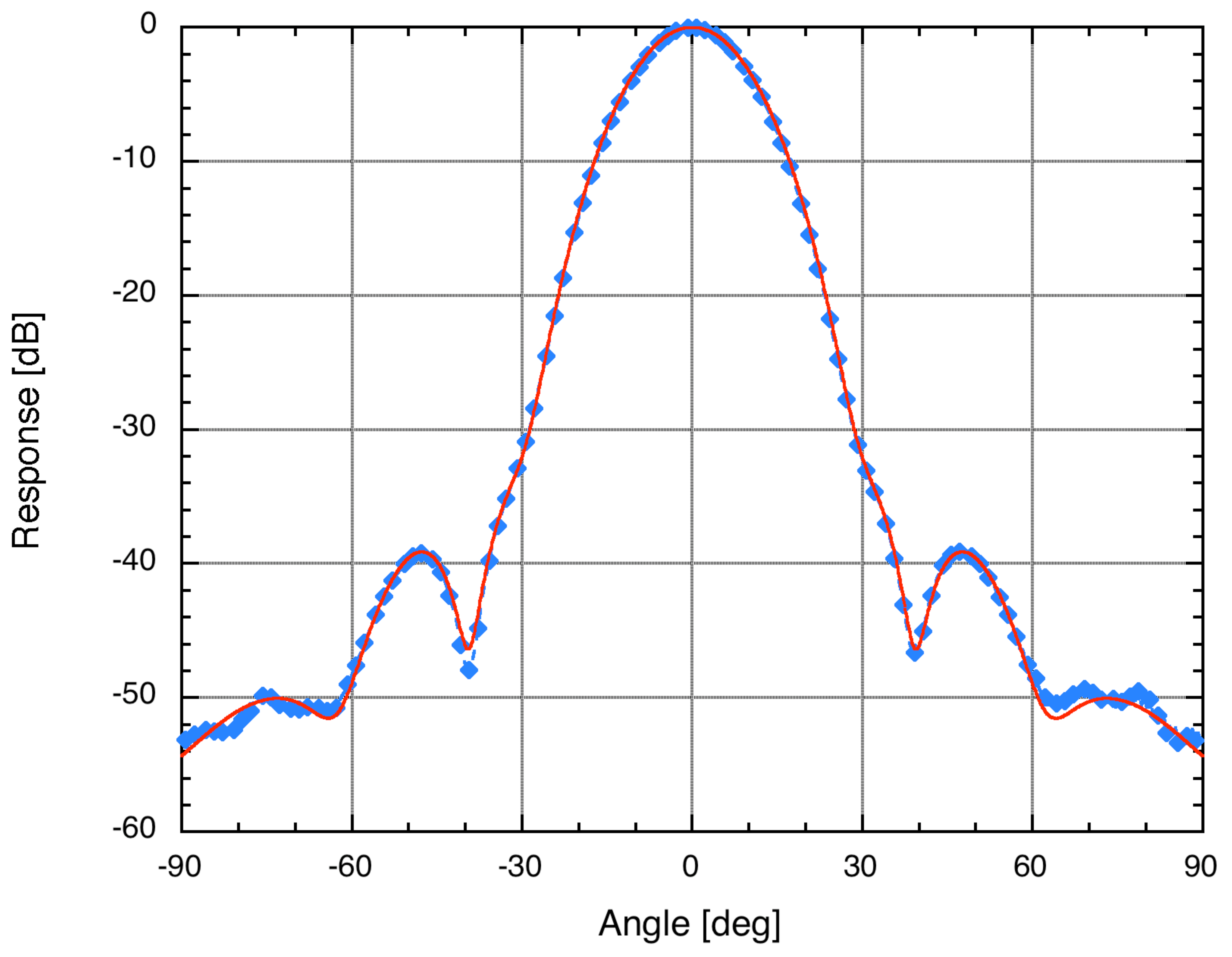}
\hfill
\includegraphics[width=0.48  \textwidth]{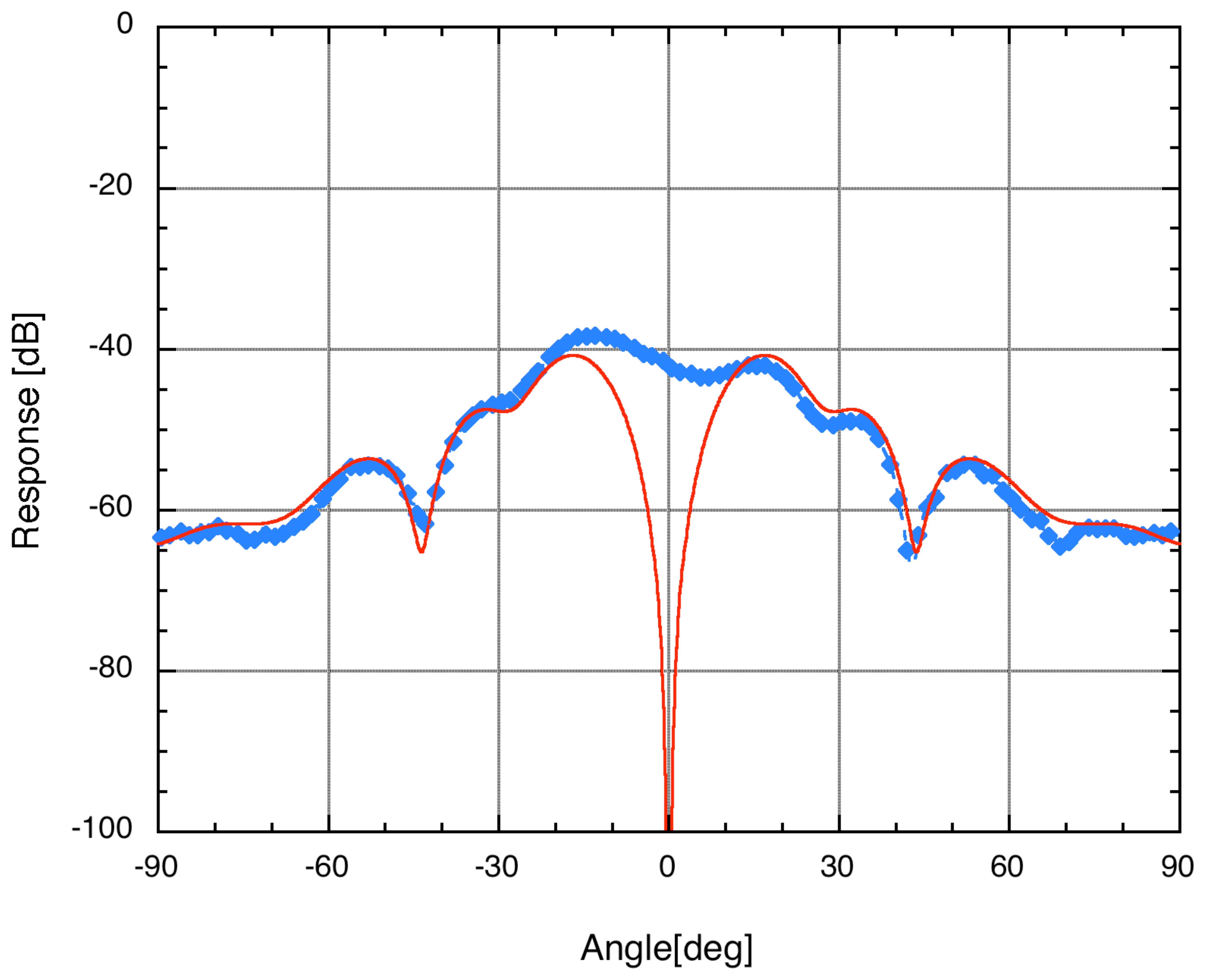}
\caption{45 degree plane co polar (left) and cross polar (right), measured (symbols) and simulated (solid line) patterns at 70 GHz.}
\label{4570}
\end{center}
\end{figure}

The amplitude patterns were measured with a horizontal scan on the angular range $\pm$90 degrees and compared with the simulated patterns, computed using the GRASP\footnote{TICRA Ltd, Copenhagen, DK, http://www.ticra.com} package in four polarization planes. Also the crosspolar component is measured  in the $\pm45^o$ planes. Typical responses are reported in figures \ref{EH30} and \ref{4530} for the 30 GHz feed horn, in figures \ref{EH44} and \ref{4544} for the 44 GHz feed horns and in figures  \ref{EH70} and \ref{4570} for the 70 GHz ones. In these figures the E, H, $45^o$ co-- and cross--polar patterns are reported for every frequency band and compared with simulations. All patterns are normalized to the maximum value, and the cross polar component is normalized to the copolar maximum. These plots show a great agreement between the measured and the simulated patterns, especially in the main lobe region. Anyway, also in the far sidelobe region, where the signal is very faint, the agreement between simulations and measurements is very good.

As already mentioned, the beam pattern was acquired both for the FH stand alone and also after the integration of the FH with the OMT. The beam pattern was measured in the E and H principal planes inserting the signal from both arms of the OMT, thus performing a scan in two orthogonal directions.

\begin{figure}[!h]
\begin{center}
\includegraphics[width=0.48 \textwidth]{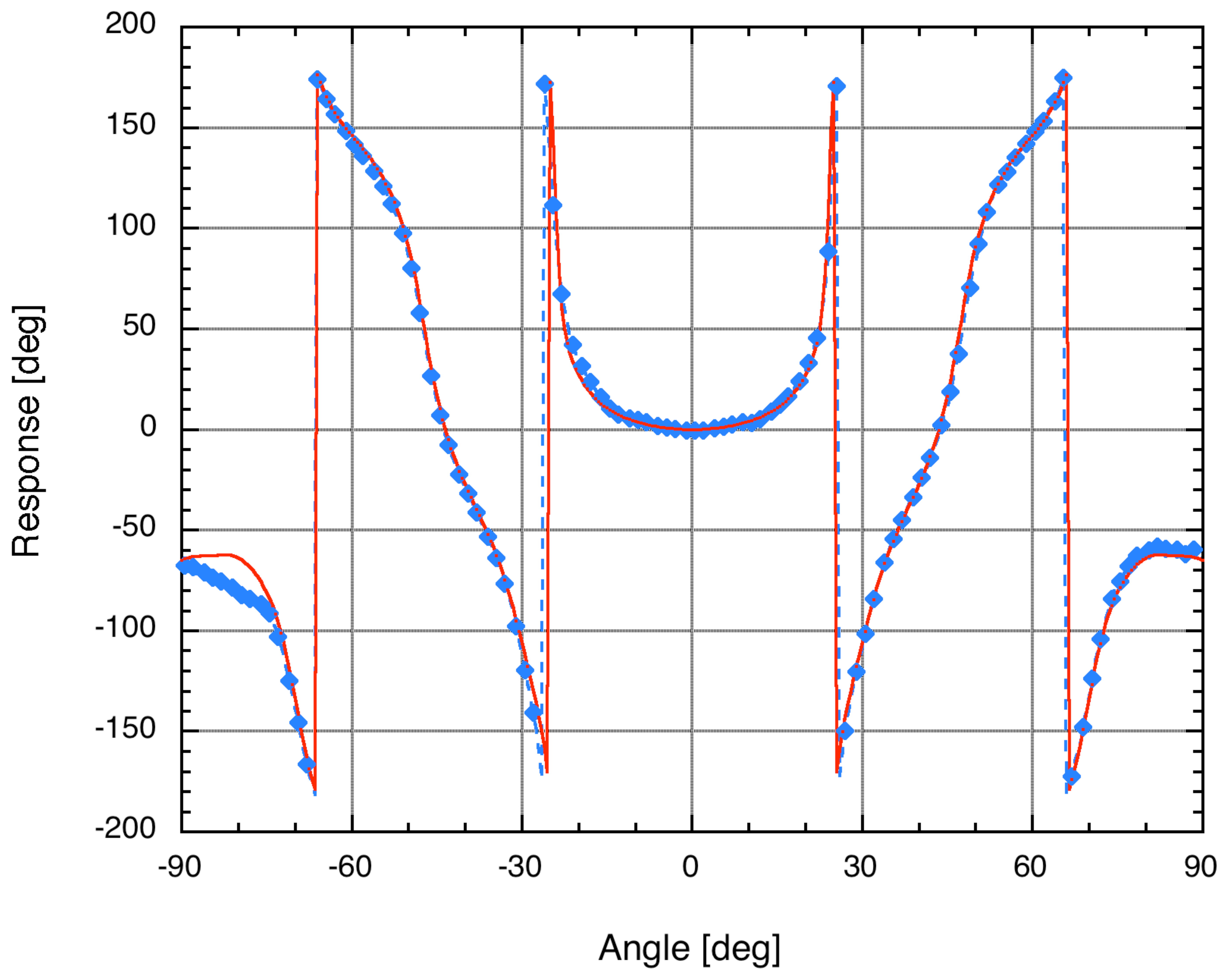}
\hfill
\includegraphics[width=0.48  \textwidth]{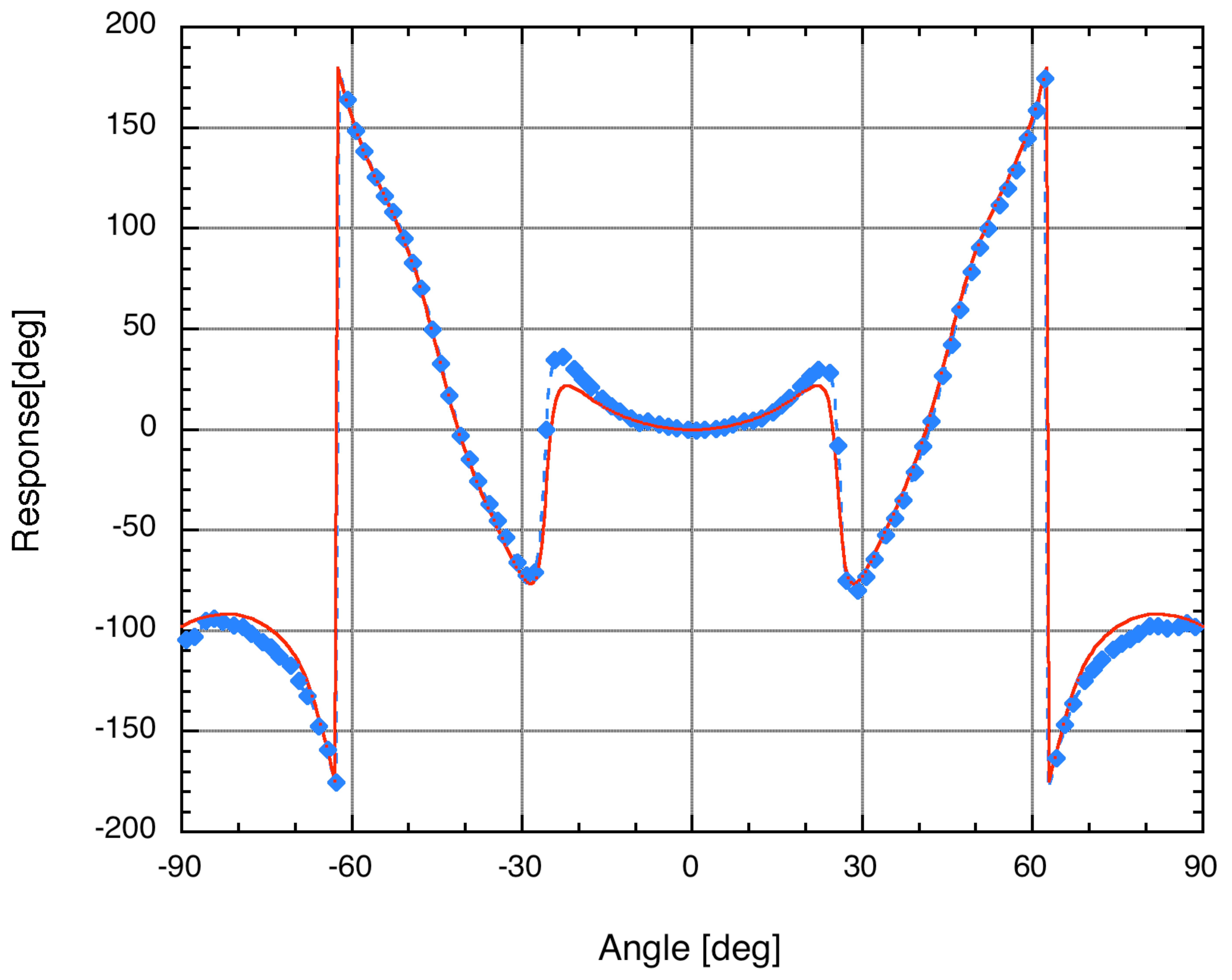}
\caption{E (left) and H (right) phase patterns measured (symbols) at 30 GHz compared with the simulation (solid line) }
\label{EH30p}
\end{center}
\end{figure}

\begin{figure}[!h]
\begin{center}
\includegraphics[width=0.48 \textwidth]{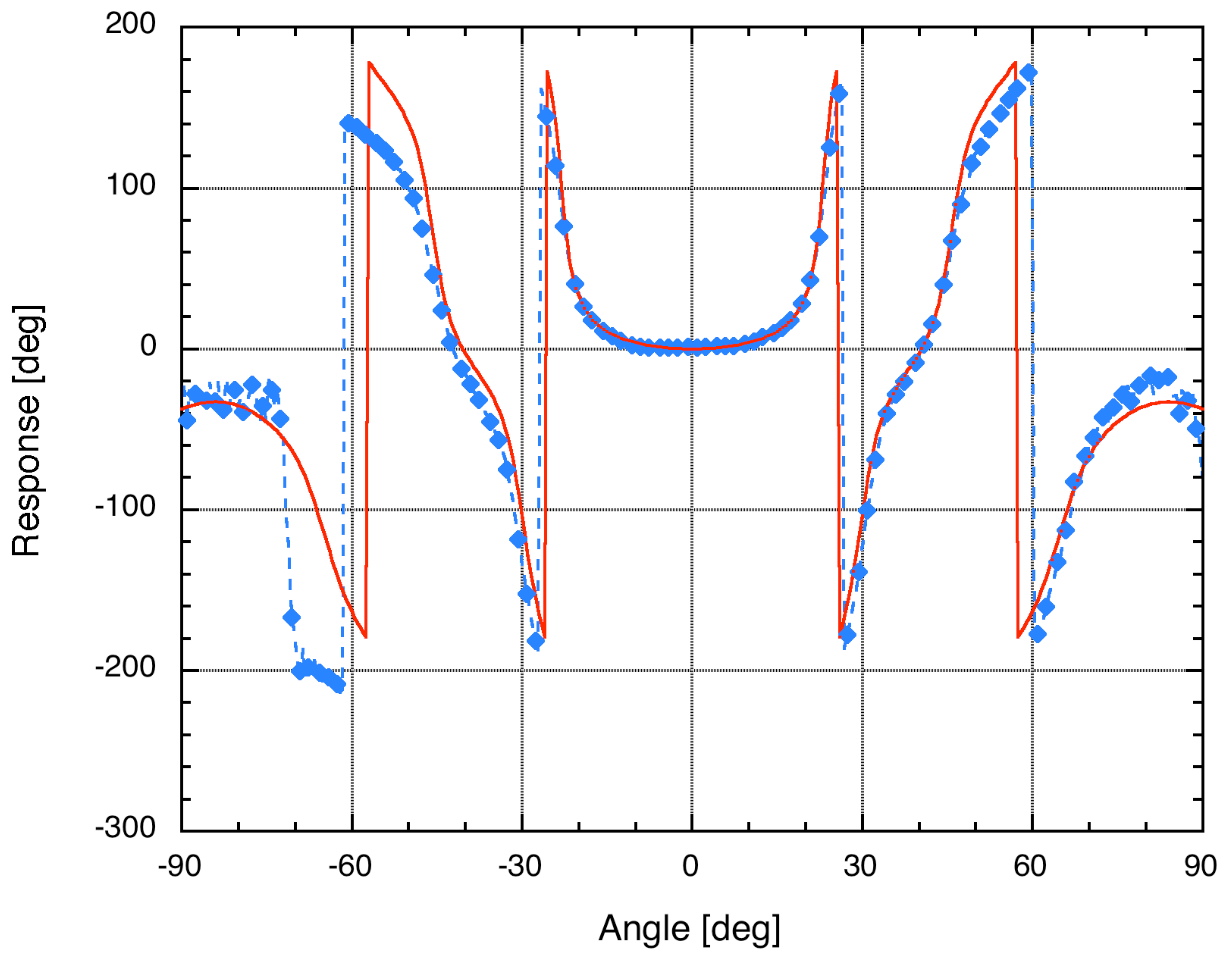}
\hfill
\includegraphics[width=0.48  \textwidth]{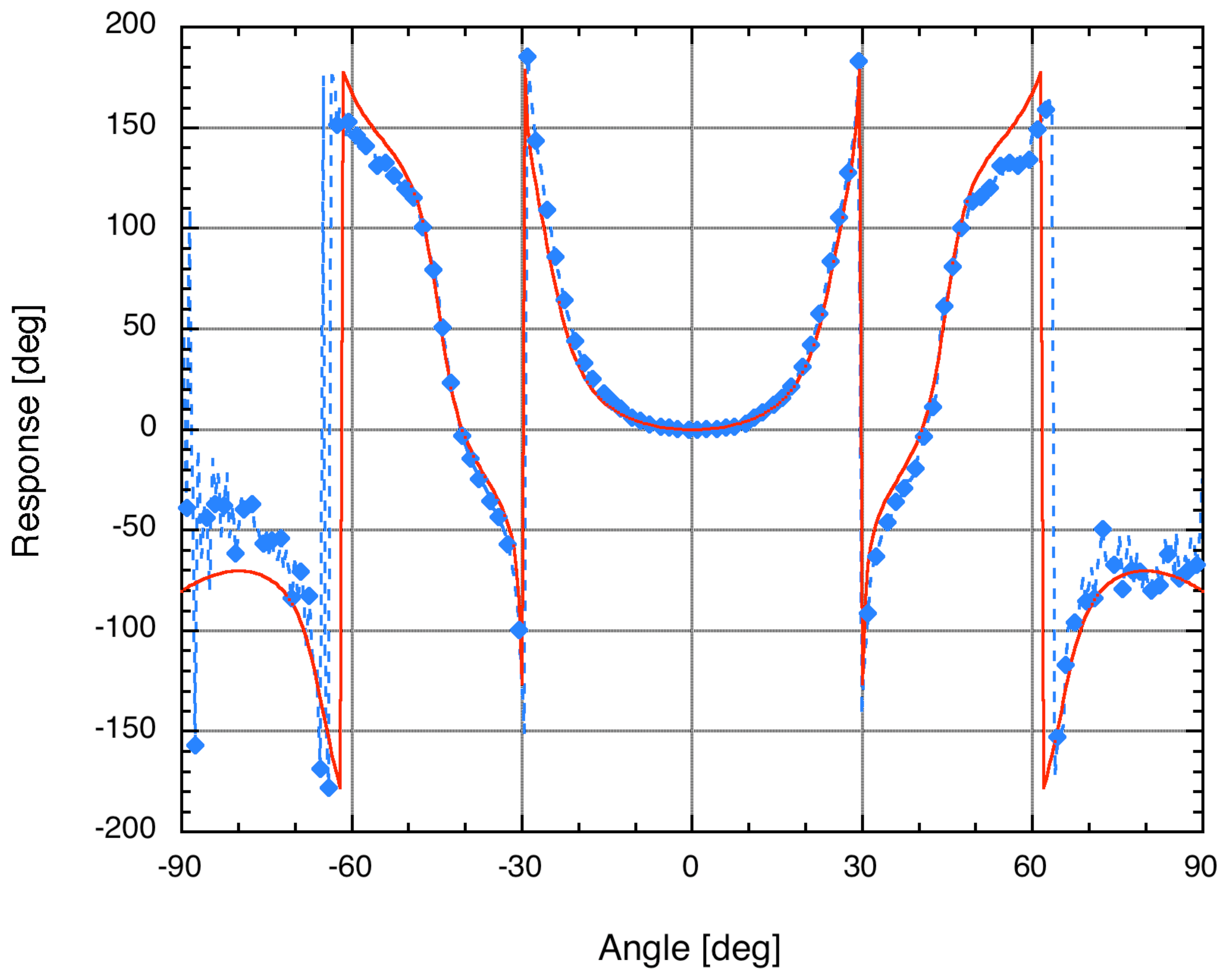}
\caption{E (left) and H (right) phase patterns measured (symbols) at 44 GHz compared with the simulation (solid line) }
\label{EH44p}
\end{center}
\end{figure}

\begin{figure}[!h]
\begin{center}
\includegraphics[width=0.48 \textwidth]{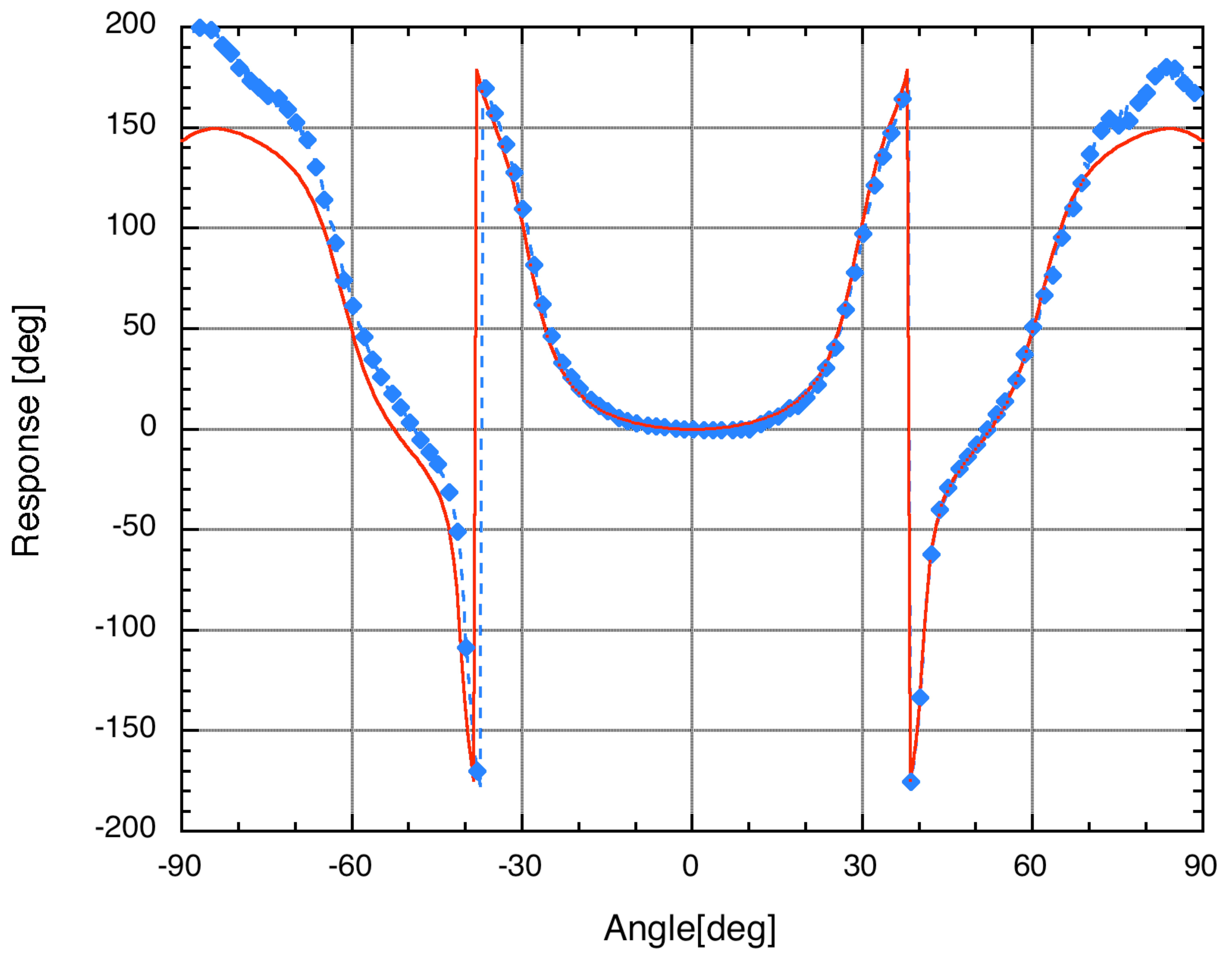}
\hfill
\includegraphics[width=0.48  \textwidth]{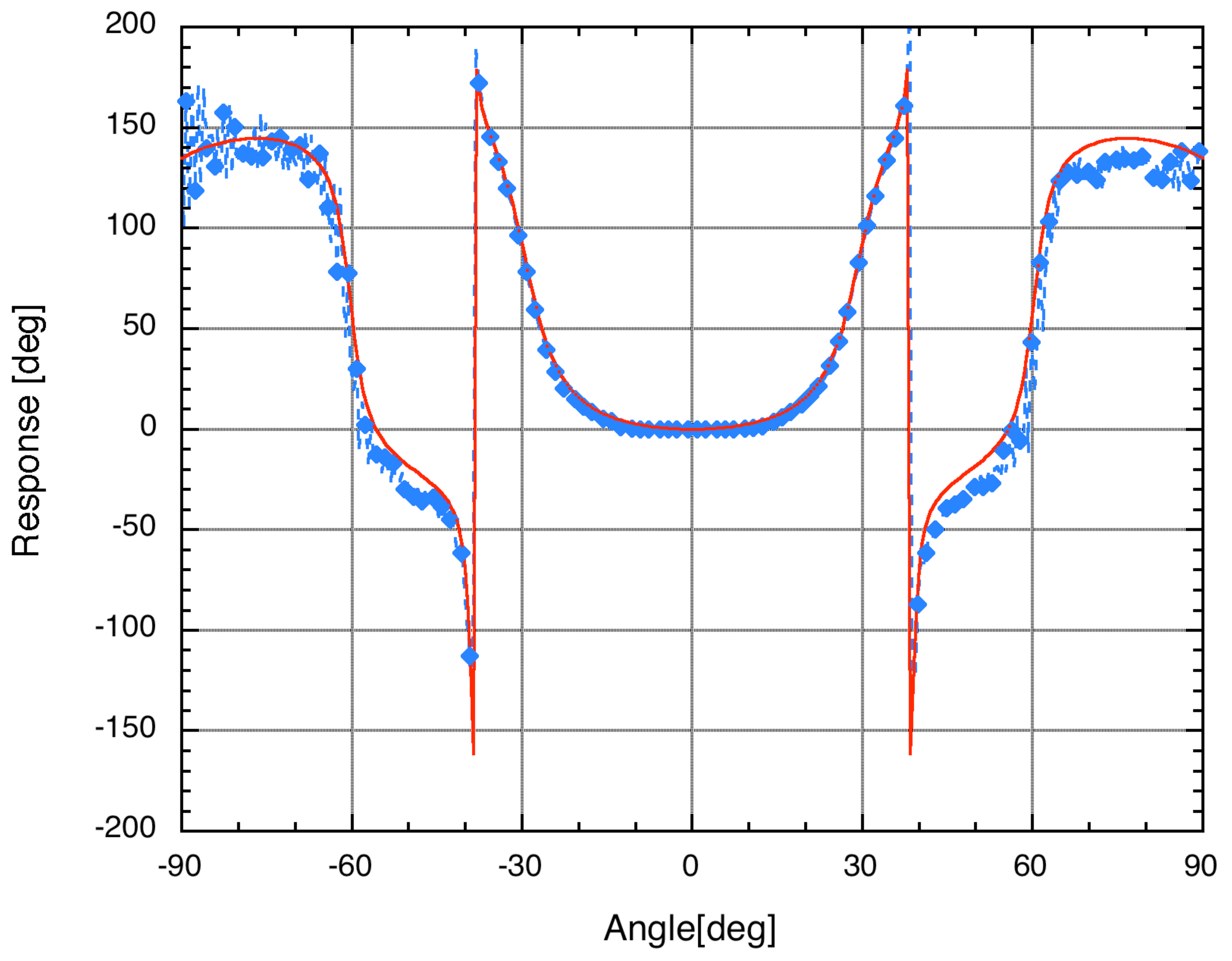}
\caption{E (left) and H (right) phase patterns measured (symbols) at 70 GHz compared with the simulation (solid line) }
\label{EH70p}
\end{center}
\end{figure}

Since all measurements were performed with a VNA, phase patterns were acquired together with the amplitude ones. Thus the phase response of the FM FHs was measured in all the planes and at all the frequencies mentioned in the previous paragraph. The agreement between measured and simulated data proves that the pattern is indeed the one designed. Examples of the results obtained measuring the phase pattern amplitude at the central frequency are reported in figures \ref{EH30p}, \ref{EH44p}, and \ref{EH70p}. 
High phase stability  cables \footnote{W.L.Gore Next Generation Microwave Cable Assembly} were used because the particular architecture  of the VNA we used makes the measurements very sensitive to that. Nevertheless, the simulated pattern reproduces the measurement to a high degree of accuracy, demonstrating the quality of both the manufacturing process and simulations.

\subsection{Return Loss}
\begin{figure}[!h]
\begin{center}
\includegraphics[width=0.8\textwidth]{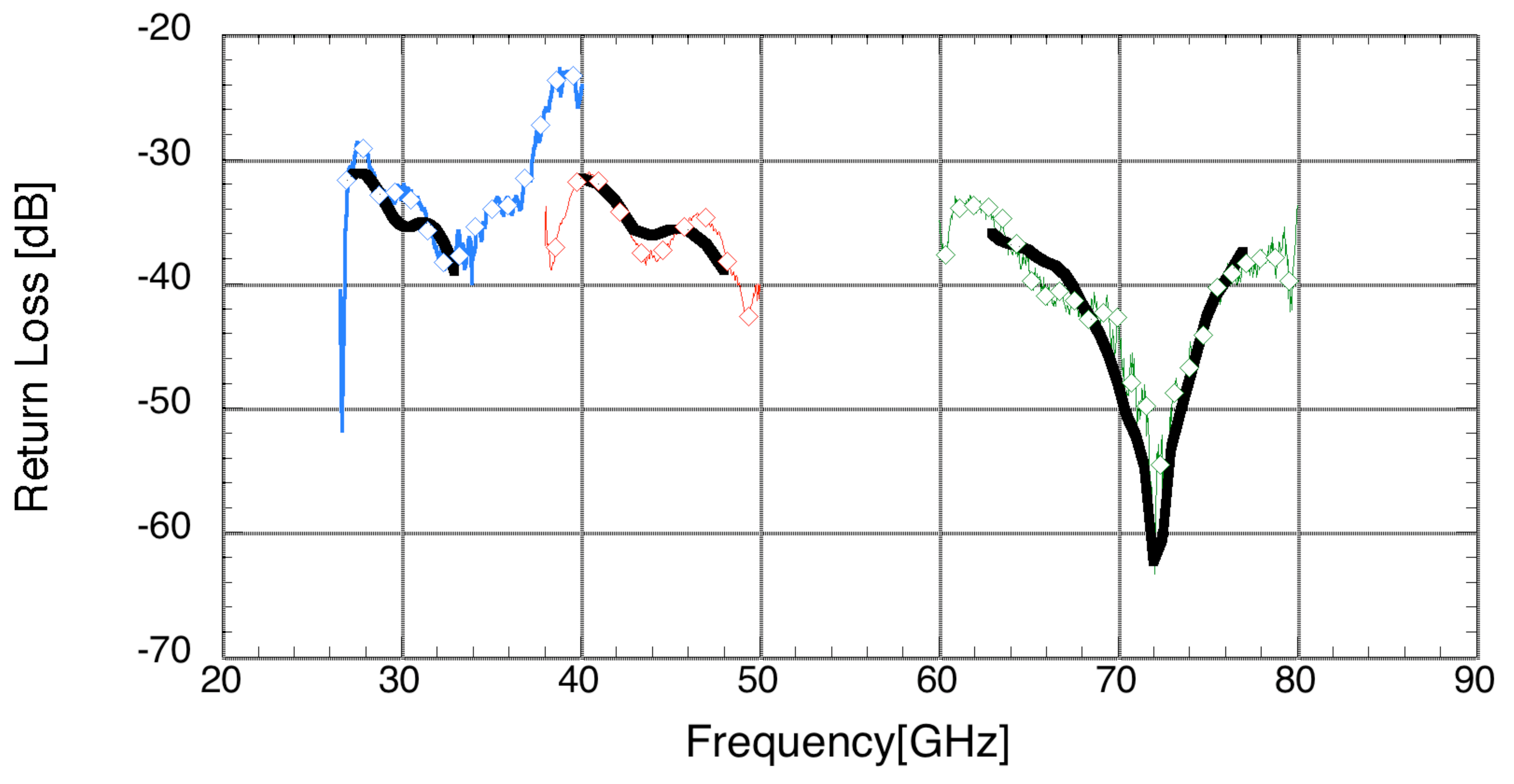}
\caption{Measured Return Loss (symbols) compared with the simulated RL (continuous line) over the three LFI frequency bands } \label{fig:RL}
\end{center}
\end{figure}

In order to obtain a full electromagnetic characterization of the FH, also the Return Loss (RL) of every single feed was measured. Of course, once integrated with the corresponding OMT, the latter dominates the reflection properties of the assembly. Nevertheless, the determination of the RL can aid in discovering possible manufacturing defects.
The RL was measured with the VNA, calibrated in a single reflectometer configuration, with the horn radiating into an Eccosorb$^{TM}$ panel; a circular to rectangular transition was used to connect the FH to the microwave circuit, but it was possible to remove its contribution to the RL signal by time domain filtering. 
Results are reported in figure \ref{fig:RL}

The RL was measured over the largest possible frequency band, but a comparison with simulation is possible only over the band of operation, where simulations are available. In this frequency range, the FHs behave exactly as expected and the signal is always below $-30$ dB level in band. 

\subsection{Vibration test}\label{vibration_test}
The LFI FHs were vibrated at space qualification level after integration with the OMTs. Vibration tests were performed with a computer--controlled  shaker, which generates the motion with the requested characteristic. In order to calibrate the fixture, an accelerometer was used to control the vibration level, while two triaxial accelerometers were used to measure the response of the unit. The assembly is submitted to sinusoidal and random vibrations and the accelerometer's recorded signals were amplified, digitized and stored for processing. The test started with a search for resonances in the range 5-2000 Hz, once the FH and OMT were firmly mounted on the shaker.

At the beginning of the test campaign, during the QM phase, the baseline procedure requested the comparison between the RL of the assembly before and after the vibration test, comparing also the Fourier Transform (FT) of the signal to localize any variations in reflection.
Because of the discovery of misalignment between the feed horn and the OMT, also the radiation pattern at the highest in--band frequency was used as a estimator of any possible misalignment due to vibration. 
During vibration tests was discovered a problem in the 70 GHz units. A macroscopic crack in the flange was observed due to the high weight of the copper feeds (the aluminum 30 and 44 GHz FH in fact passed the vibration test easily). All the flanges were reinforced and the damaged units replaced with spare units. 

\section{conclusions}\label{chapt:conclusions}
Eleven dual profiled corrugated feed horns were developed, manufactured, and tested as flight units of the Low Frequency Instrument onboard the ESA Planck satellite. The feed horns were based on six different electromagnetic designs obtained as a result of a long iteration process that involved optical calculations and mechanical constraints on the focal plane unit. The qualification campaign, mainly focused on RF return loss and pattern (amplitude and phase) measurements, was successfully concluded and the misalignment problems were solved by integrating the feeds with the OMT with an optical endoscope and measuring the pattern at the highest in--band frequency. Vibration test were performed at integrated level (feed/OMT) showing a problem with the 70 GHz units. It was solved by reinforcing the flanges to prevent any further cracks as additional vibration tests demonstrated. 
The agreement between the pattern measurements and the expected performances (simulated using nominal corrugation profile) is excellent both in amplitude and in phase. Moreover reflection measurements show a good impedance match for all the horns, the return loss being better than -$30$ dB over the whole $20\%$ of operational bandwith.

\acknowledgments
Planck is a project of the European Space Agency with instruments funded by ESA member states, and with special contributions from Denmark and NASA (USA). The Planck-LFI project is developed by an International Consortium lead by Italy and involving Canada, Finland, Germany, Norway, Spain, Switzerland, UK, USA. The Italian contribution to Planck is supported by the Italian Space Agency (ASI). 

\end{document}